# Octave-wide broadening of ultraviolet dispersive wave driven by soliton-splitting dynamics

Tiandao Chen[1,4,5,†], Jinyu Pan[1,3,4,†], Zhiyuan Huang[1,2,*], Yue Yu[1,3,4], Donghan Liu[1,2,4], Xinshuo Chang[1,3,4], Zhengzheng Liu[1], Wenbin He[2], Xin Jiang[2], Meng Pang[1,2,3,*], Yuxin Leng[1,3,*] and Ruxin Li[1,5]

[1]State Key Laboratory of High Field Laser Physics and CAS Center for Excellence in Ultra-intense Laser Science, Shanghai Institute of Optics and Fine Mechanics (SIOM), Chinese Academy of Sciences (CAS), Shanghai 201800, China

[2]Russell Centre for Advanced Lightwave Science, Shanghai Institute of Optics and Fine Mechanics and Hangzhou Institute of Optics and Fine Mechanics, Hangzhou, 311421, China

[3]Hangzhou Institute for Advanced Study, University of Chinese Academy of Sciences, Hangzhou 310024, China

[4]Center of Materials Science and Optoelectronics Engineering, University of Chinese Academy of Sciences, Beijing 100049, China

[5]Zhangjiang Laboratory, Shanghai 201210, China

*Corresponding author: huangzhiyuan@siom.ac.cn; pangmeng@siom.ac.cn; lengyuxin@siom.ac.cn

†These authors contribute equally to the work

**Coherent dispersive wave emission, as an important phenomenon of soliton dynamics, manifests itself in multiple platforms of nonlinear optics from fibre waveguides to integrated photonics. Limited by its resonance nature, efficient generation of coherent dispersive wave with ultra-broad bandwidth has, however, proved difficult to realize. Here, we unveil a new regime of soliton dynamics in which the dispersive wave emission process strongly couples with the splitting dynamics of the driving pulse. High-order dispersion and self-steepening effects, accumulated over soliton self-compression, break the system symmetry, giving rise to high-efficiency generation of coherent dispersive wave in the ultraviolet region. Simultaneously, asymmetric soliton splitting results in the appearance of a temporally-delayed ultrashort pulse with high intensity, overlapping and copropagating with the dispersive wave pulse. Intense cross-phase modulations lead to octave-wide broadening of the dispersive wave spectrum, covering 200 to 400 nm wavelengths. The highly-coherent, octave-wide ultraviolet spectrum, generated from the simple capillary fibre set-up, is in great demand for time-resolved spectroscopy, ultrafast electron microscopy and frequency metrology applications, and the critical role of the secondary pulse in this process reveals some new opportunities for all-optical control of versatile soliton dynamics.**

## Introduction

Temporal solitons in optical waveguides, originated from dynamical balancing between Kerr nonlinearity and chromatic dispersion[1-3], underlie a rich variety of phenomena in nonlinear optics. Some key examples include ultrashort pulse self-compression[4-6], efficient light conversion over a large wavelength range[6-11], and multi-octave-spanning supercontinuum generation[12-17]. Among them, the phenomenon of resonant dispersive wave (DW) emission[6,10-14,18-23], also known as optical Cherenkov radiation[22], has been at the focus of attention across multiple platforms of nonlinear optics, from optical fibre waveguides[6,10-14] to integrated photonics chips[16,24-27]. Through introducing higher-order (>2$^{nd}$ order) dispersion effects, the otherwise symmetric dispersion landscape of an optical waveguide is modulated, leading to the appearance of additional zero group-velocity-dispersion (GVD) points. This modulation also gives birth to phase-matched wavelengths in the normal GVD regime of the waveguide, coherently linked to the soliton (pump) wavelength in the anomalous regime[6,10-14,18-23]. In such an optical waveguide, spectral spreading of a high-order soliton over propagation could enable, through the cascaded four-wave mixing[23], efficient energy conversion from the pump pulse to a packet of quasi-linear optical waves at the phase-matched wavelength. This phenomenon of coherent DW emission has facilitated the generation of octave-spanning supercontinuum[12-16,27], which is a critical pre-requisite of the f-2f self-referencing technique for optical frequency comb generation[28-30]. Moreover, coherent DW emission has also been investigated in a variety of nonlinear optical systems based on multi-mode fibres[31-34], hollow-core fibres[6,10-14,35-42] and integrated silica or silica nitride waveguides[16,43,44], highlighting the importance of this nonlinear laser-frequency-conversion mechanism for extending the wavelength accessibility of ultrafast light sources, towards both ultraviolet (UV)[6,10-13,37,38] and mid-infrared[16,36,45].

Because of the intrinsic nature of phase matching[23], the manifestation of DW emission in these aforementioned systems is commonly observed as resonant spikes on the soliton spectrum[6,10]. This narrow-band feature of DW emission restricts, to some extent, the spectral coverage and therefore the ultrafast performance of DW pulses. Some attempts for broadening DW spectra, based on some mechanisms such as soliton-trapping[46,47], Raman nonlinearity[12] or photoionization effect[13], have been reported in literatures. Yet, the realization of coherent DW emission with simultaneously high conversion efficiency and ultra-broad spectral coverage, supporting the formation of high-energy, single-cycle or even sub-cycle pulses, remains to be a long-standing challenge for many years.

In this article, we present a unique regime of nonlinear soliton dynamics, obtained in a gas-filled hollow capillary fibre (HCF), that involves a delicate coupling between UV DW emission and soliton splitting effects. In the nonlinear dynamics, the self-steepening-induced blue shift of the pump pulse, accumulated over the soliton self-compression process, enhances the radiation efficiency of the short-wavelength DW[10,48]. Then, the generated UV DW pulse, after a short length of propagation in the capillary fibre, interacts intensely with a secondary pulse which stems from the asymmetric splitting process of the pump pulse. With a few-femtosecond pulse width, the secondary pulse has a high peak intensity comparable to that of the self-compressed main pulse, strongly modulating the UV DW. As shown in the following, this delicate coupling leads to spectral broadening of the DW pulse by nearly one order of magnitude, giving rise to the generation of octave-wide UV spectrum in this capillary fibre set-up, with good spectral stability, high phase coherence and some wavelength tunability. This work extends our knowledge on interactions between soliton and DW, and further highlights the versatility of nonlinear soliton dynamics in optical waveguides.

## Broadband ultraviolet DW generation

The gas-filled capillary fibre set-up, used to perform the experiments, is sketched in Fig. 1a (see Methods below and Section 1 of the Supplementary Information for more details of the set-up). The silica-glass capillary has a length of 42 cm and an inner diameter of 100 μm. When filled with monatomic (Raman inactive) gas of Ne, the dispersion landscape of this capillary waveguide can be adjusted through varying the gas pressure[6,7,10,49,50]. At a gas pressure of 5 bar, the zero GVD point of the waveguide is at 520 nm. In this condition, the waveguide exhibits an anomalous GVD value of $-4.39 \times 10^{-2}$ ps$^2$/km at the pump wavelength (800 nm), while the phase-matching wavelength of the DW emission[50] (when neglecting the nonlinearity term in the formula) can be calculated to be 255 nm, see Fig. 1a.

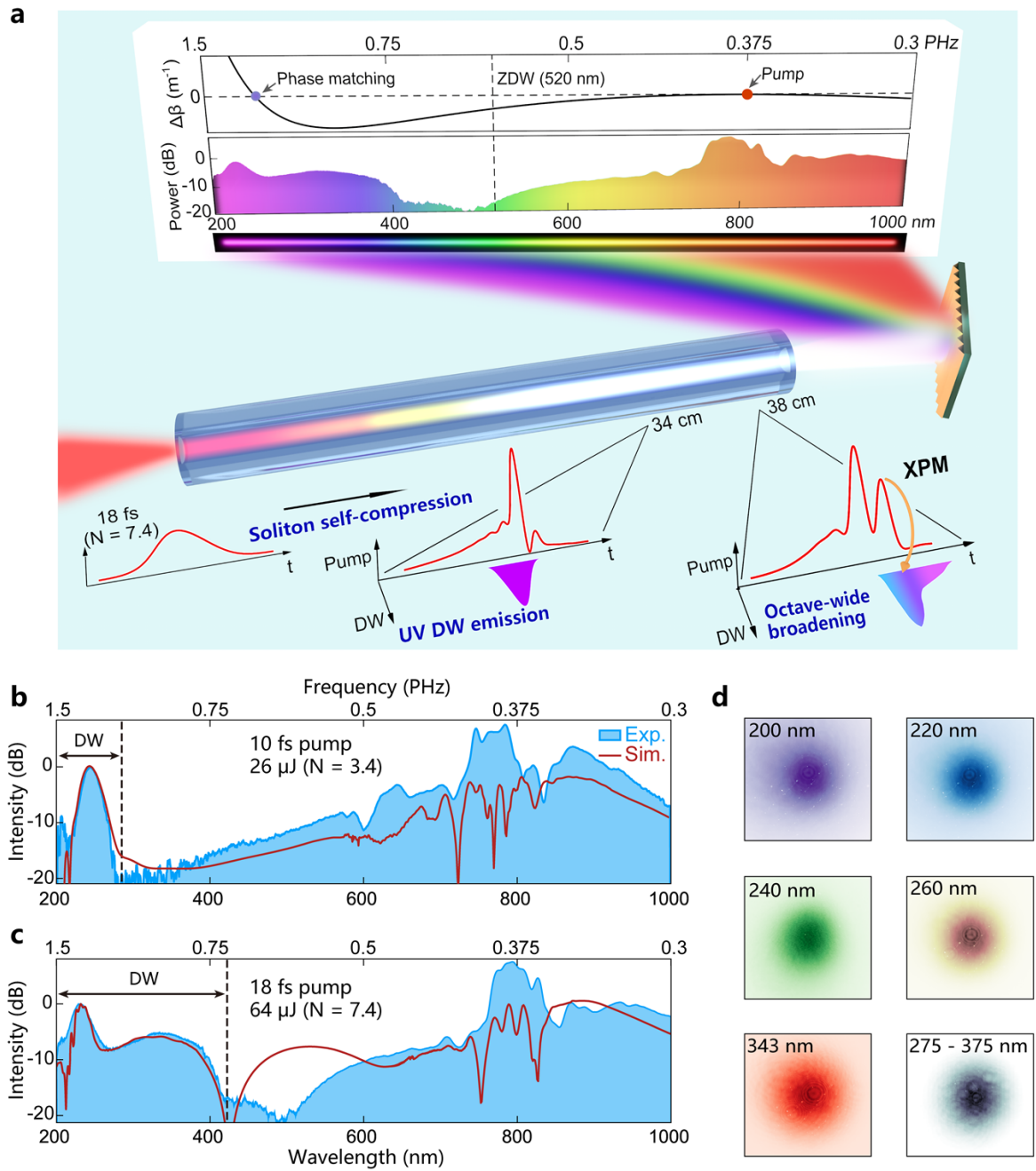

**Fig. 1 | Sketch of the broadband UV DW generation mechanism, experimental and numerical simulation results. a**, Broadband UV DW generation mechanism in Ne-filled HCF. At a gas pressure of 5 bar, the zero-dispersion wavelength (ZDW) is 520 nm. The dynamics of soliton self-compression, UV DW emission, cross-phase modulation (XPM) caused by soliton-splitting effect, and octave-wide broadening of DW spectrum are illustrated at the bottom. **b,c**, Measured and simulated spectra when pumped at 10 fs, 26 μJ (N = 3.4) (**b**) and 18 fs, 64 μJ (N = 7.4) (**c**). The HCF used in the experiment has a length of 42 cm and a core diameter of 100 μm and filled with 5 bar Ne gas. **d**. Near-field beam profiles of UV DW at different wavelengths, measured using a CCD camera.

In the experiment, we obtained significantly-different DW spectra (see Methods below) at the output port of the capillary fibre, when using different pump pulse parameters, as illustrated in Figs. 1b,c. When the 10-fs pump pulse with an energy of 26 µJ (corresponding to a soliton order of 3.4) was used, typical resonance-shaped DW spectrum was observed with a spectral bandwidth of ~20 nm (see Fig. 1b). In this case, the energy fraction of the UV DW energy in the output spectrum is measured to be ~6.8%. A much broader DW spectrum, covering the entire 200 to 400 nm UV region (within 10 dB spectral density scale), could be obtained (see Fig. 1c) as we altered the pump pulse parameters to 18 fs and 64 µJ, corresponding to a higher soliton order of 7.4. This octave-spanning UV DW, measured at the fibre output port, has an energy fraction of ~15.5% and exhibits perfect single-fundamental-mode purity over its entire spectral range. Near-field beam profiles of the broadband DW were measured at several wavelengths, using narrow-band optical filters and an UV charge-coupled-device (CCD) camera, see Fig. 1d (see also Methods below).

These experimental results of both narrow- and broad-band UV DW emission processes can be quantitively reproduced using both the generalized nonlinear Schrödinger equation (GNLSE)[51] or the single-mode unidirectional pulse propagation equation (UPPE)[52-54], see Figs. 1b,c (see also Methods below). We found that the numerical results of nonlinear pulse propagation processes in the capillary fibre, simulated using these two models, exhibit excellent agreement with each other, verifying the validity of the GNLSE model in the few-cycle regime[51,55]. See Section 2 of the Supplementary Information for detailed discussions. The system parameters used in the simulations were exactly the same as these in the experiments (see Fig. 1), and the temporal profile of the pump pulse used in the simulation comes from the experimental results measured using the frequency-resolved optical gating (FROG) set-up, see more details in Section 4 of the Supplementary Information. The GNLSE used in the simulation can be expressed as

$$\frac{\partial A}{\partial z} + \frac{i\beta_2}{2}\frac{\partial^2 A}{\partial T^2} - i\gamma |A|^2 A = \sum_{n=3}^{\infty} i^{n+1}\frac{\beta_n}{n!}\frac{\partial^n A}{\partial T^n} - \frac{\gamma}{\omega_0}\frac{\partial}{\partial T}(|A|^2 A) - \frac{\alpha}{2}A + P_{ion} \quad (1)$$

where $A$ is the complex amplitude of the light field in time domain, $\beta_n$ the n$^{th}$-order dispersion of the fibre waveguide, $T$ the time in a reference frame moving at the group velocity $1/\beta_1$, $\gamma$ the nonlinear parameter, $\omega_0$ the central frequency of the driving pulse, $\alpha$ the wavelength-dependent fibre loss, and $P_{ion}$ the photoionization-related term. In the left side of Eq. (1), the Kerr nonlinearity and 2$^{nd}$-order dispersion terms describe the propagation of canonical solitons, while in the right side of the equation, four terms are included, describing respectively higher-order dispersion, self-steepening effect, wavelength-dependent capillary loss and photoionization effect. All of these four terms could be regarded as some perturbations to propagation processes of canonical solitons which (without perturbations) exhibit perfect symmetry in both spectral and temporal domains[51,56].

As conceptually illustrated in Fig. 1a, the first stage of propagation of the pump pulse in the capillary roughly follows the high-order soliton propagation process governed by the nonlinear Schrödinger equation[1,51,56], leading to soliton self-compression. This is because at this beginning stage, the relatively-narrow pulse spectrum (long pulse duration) leads to relatively-weak high-order effects. However, the soliton self-compression results in significant broadening of the pulse spectrum and increased pulse peak power, enhancing higher-order effects as described in the right side of Eq. (1). The first obvious consequence resulting from these high-order effects is the emission of phase-matched UV DW[6,10-14,18-23], which can be further enhanced through blue shifting of the pump pulse due to self-steepening nonlinearity[10,48]. The DW has generally a narrow-band spectrum, limited by the phase matching nature of its emission process[23]. After being generated, the DW pulse has a lower group velocity than the soliton, and therefore moves gradually behind the pump pulse.

Near the maximum self-compression point, strong high-order effects (perturbations) could also cause the highly-asymmetric splitting of the high-order soliton[18,19,57,58], leading to the appearance of the secondary pulse just behind the main pulse (with a short temporal delay), see Fig. 1a. In principle, we could adjust the relative intensity and temporal position of this secondary pulse through varying the pulse width and soliton order of the pump pulse[18,19,57,58], so as to enable perfect overlapping and strong interactions between the secondary pulse and the DW. Significant spectral broadening of the UV DW could therefore be obtained through intense cross-phase modulations, see Fig. 1a.

## Spectrogram analysis

To better illustrate the complex soliton dynamics, we performed deliberately numerical studies based on the spectrogram-analysis method. The simulation results and our analyses are shown in Fig. 2, see also Supplementary Movie 1 and Movie 2. While the simulated evolution dynamics of the 18-fs, 64-μJ (soliton order N = 7.4) pump pulse inside the capillary fibre are illustrated in the right column of Fig. 2 (Figs. 2e-h), the evolution dynamics of the 10-fs, 26-μJ pulse (soliton order N = 3.4) in the capillary fibre are also illustrated in the left column (Figs. 2a-d) for a direct comparison. In the simulation, the temporal profiles of the pump pulse (plotted as red lines in Fig. 2) are obtained through using an ideal (chirp-free) square-shaped filter with a transmission window covering 400 to 4000 nm, while the temporal profiles of the DW pulse (plotted as blue shadows) are obtained using another ideal filter covering 100 to 400 nm. It can be found that at a moderate soliton order of 3.4 (left column of Fig. 2), soliton self-compression results in a strong spectral broadening of the pump pulse. Near the maximum compression point ($L = 28$ cm, see Fig. 2b), the phenomenon of narrow-band DW emission is clearly observed at the phase-matching wavelength (marked with **I** in Fig. 2). After this maximum compression point, the peak power of the pump pulse decreases gradually over propagation (see Figs. 2c,d), without observing soliton splitting phenomenon. The generated DW pulse gradually moves behind the pump pulse due to its lower group velocity, and its temporal stretching is due to normal dispersion of the capillary waveguide at this phase-matching wavelength.

As illustrated in the right column of Fig. 2, a higher soliton order of 7.4 can lead to a dramatic change of the nonlinear pulse-evolution dynamics, especially near and after the maximum compression point ($L \geq 34$ cm, see Figs. 2f-h). The higher peak power of the compressed pulse results in stronger higher-order nonlinear effects (in the right side of Eq. (1)) accumulated over the soliton self-compression process, which, through collaborating with higher-order dispersion effects, leads to highly-asymmetric soliton splitting phenomenon[18,19] (marked with **II** in Fig. 2). See more detailed results and discussions in Section 3 of the Supplementary Information. It can be found in Fig. 2f that the asymmetric soliton splitting gives birth to a secondary pulse which, in the simulation, is around 4 fs behind the main pulse. As illustrated in Fig. 2g, the intensity of this secondary pulse increases gradually over propagation, and its peak power reaches to ~1.6 GW at $L = 38$ cm where it has the largest overlapping with the UV DW pulse. Strong cross-phase modulations cause obvious spectral broadening of the UV DW[51,59,60], marked with **III** in Fig. 2h. The leading edge of DW pulse moves to the longer wavelengths (red shift) due to a positive slope of the index modulation, while light components at the trailing edge of the DW pulse moves to the shorter wavelength (blue shift) due to a negative slope.

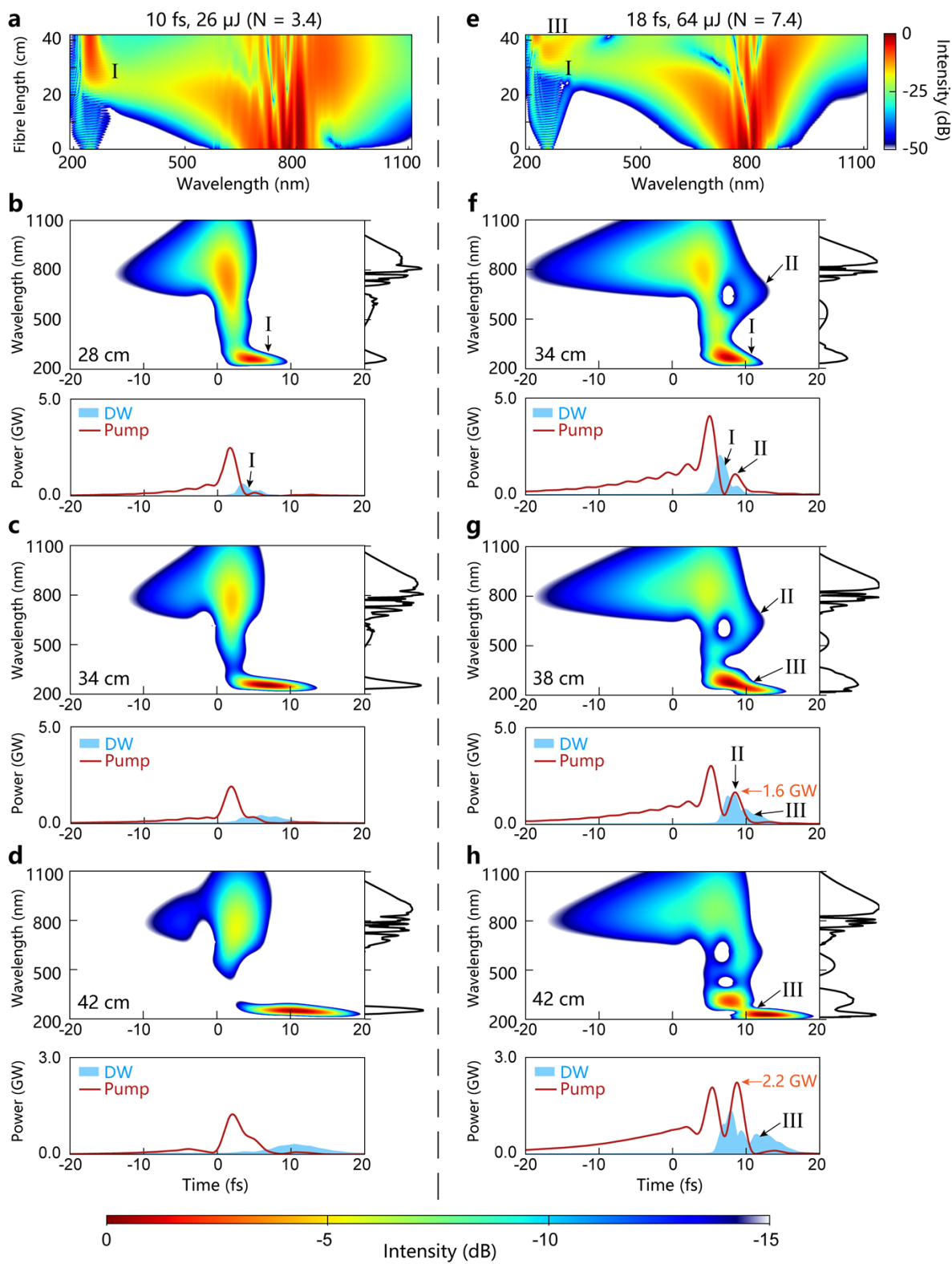

**Fig. 2 | Spectral evolution and spectrogram analysis of the pulse propagating in the HCF in Figs. 1b,c.** Simulated evolution dynamics of the 10 fs, 26 μJ pulse (N = 3.4) (**a-d**) and the 18 fs, 64 μJ pulse (N = 7.4) (**e-h**), with the same simulation parameters as Figs. 1b,c. **a,e,** Simulated spectral evolution of the pulse as a function of fibre length. **b-d,f-h,** Spectrogram analysis of the pulse at the indicated positions in the HCF. **I** – Narrow-band DW emission, **II** – asymmetric soliton splitting phenomenon, **III** – spectral broadening of the UV DW through cross-phase modulation. Through using the ideal filters with transmission windows covering 400 to 4000 nm and 100 to 400 nm, the temporal profiles of the pump pulse and DW were

obtained, respectively, which were plotted as red lines and blue shadows. The black lines indicate the pulse spectra at different fibre positions.

As illustrated in Fig. 2h, at the trailing edge of the DW pulse, the blue-shifting components have even lower group velocities than the pump pulse, thus quickly moving away from the secondary pulse. The increased walking-off effect restricts further broadening of the DW spectrum towards even shorter wavelengths. On the contrary, the red-shifting components at the leading edge accelerate due to normal waveguide dispersion, which strengthens the interaction between the two pulses and therefore enhances the red-shifting effect of the DW spectrum. This mechanism, concerning acceleration and deceleration of different DW components, can be used to explain some critical spectral features (observed in the experiments) of this broadband DW, including the formation of M-shaped spectrum and its obvious spectral asymmetry biased to the longer wavelengths (see Fig. 1c and Fig. 3).

Note that in these numerical simulations, we found that the peak intensity of the pump pulses is always far below the photoionization threshold of the Ne gas in the capillary fibre. Therefore, the photoionization (4$^{th}$) term in the right side of Eq. (1) can be neglected without inducing any influence on the simulated results. We also found in the simulation that slight change of the pump pulse shape has little impact on its nonlinear propagation dynamics. For example, we performed numerical simulations using a perfect Gaussian-shaped pump pulse with the same energy and similar temporal width as shown in Fig. 1c, and in this case the simulated broadband DW generation process is very similar as that shown in Figs. 2e-h. See Section 3 of the Supplementary Information and Supplementary Movie 3.

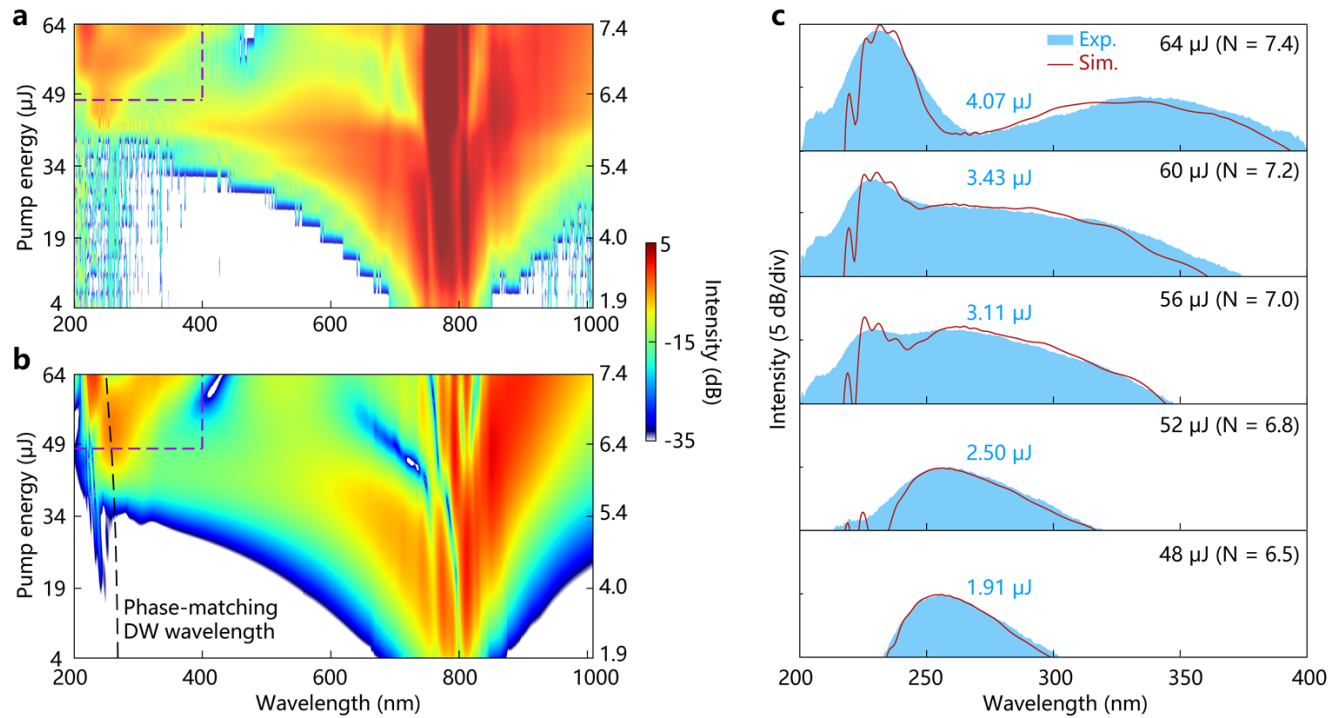

**Fig. 3 | Experimental and numerical results of DW spectral bandwidth adjustments. a**, Experimental spectral evolution at the output of the 42-cm-long HCF with 100 μm core diameter filled with 5 bar Ne gas for 18 fs pump pulses as a function of pump pulse energy. **b**, Corresponding numerical results. The black dashed line indicates the phase-matching DW wavelength with considering the nonlinearity term. **c**, Several examples of experimental (blue shadows) and simulated (red lines) spectral evolution of DW at pump pulse energies from 48 μJ to 64 μJ, corresponding to the purple dashed frames in (**a**,**b**). The results in (**c**) are also plotted on a liner scale, see Supplementary Fig. S15. The DW spectral bandwidth adjustments at 6 bar and 7 bar Ne gas are shown in Supplementary Fig. S12.

## Adjustment of UV DW spectrum

The spectral bandwidth of the UV DW can be adjusted simply through varying the pump pulse energy launched into the capillary fibre. In the experiment, we gradually decreased the energy of 18-fs pump pulse from 64 µJ to 4 µJ, and measured the output spectrum at the fibre output port. The experimental results are illustrated in Fig. 3a. For a comparison, numerical simulations are illustrated in Fig. 3b, and excellent agreement is found between the experimental and numerical results. Several typical examples at pulse energies from 48 µJ to 64 µJ, corresponding to soliton orders from 6.5 to 7.4, are plotted in Fig. 3c. It can be found that at a higher soliton order, stronger broadening effect of the UV DW spectrum could be obtained, which is mainly attributed to an increased energy (and therefore a higher peak power) of the secondary pulse generated in the soliton splitting process.

## Coherence and stability characterizations

The phase profile and coherence performance of the generated DW were investigated numerically. As illustrated in Fig. 4a, the broadband DW exhibits a smooth spectral phase profile at the output port of the fibre. We also simulated the influence of quantum noise on the broadband DW emission process, and estimated the degree of first-order coherence[13,56,61,62] on the broadband DW spectrum (see Methods below). The results (see Fig. 4a) show that the generated DW has good coherence over its broad spectrum. In the simulation, we found that the chirp of the output DW pulse (see Fig. 4b) could be well compensated through introducing $-2$ fs$^2$ group delay dispersion (GDD) to the pulse. As illustrated in Fig. 4c, the compressed pulse has a full-width-half-maximum (FWHM) width of 1.3 fs, which is quite close to its Fourier-transform-limit (FTL) value (1.2 fs).

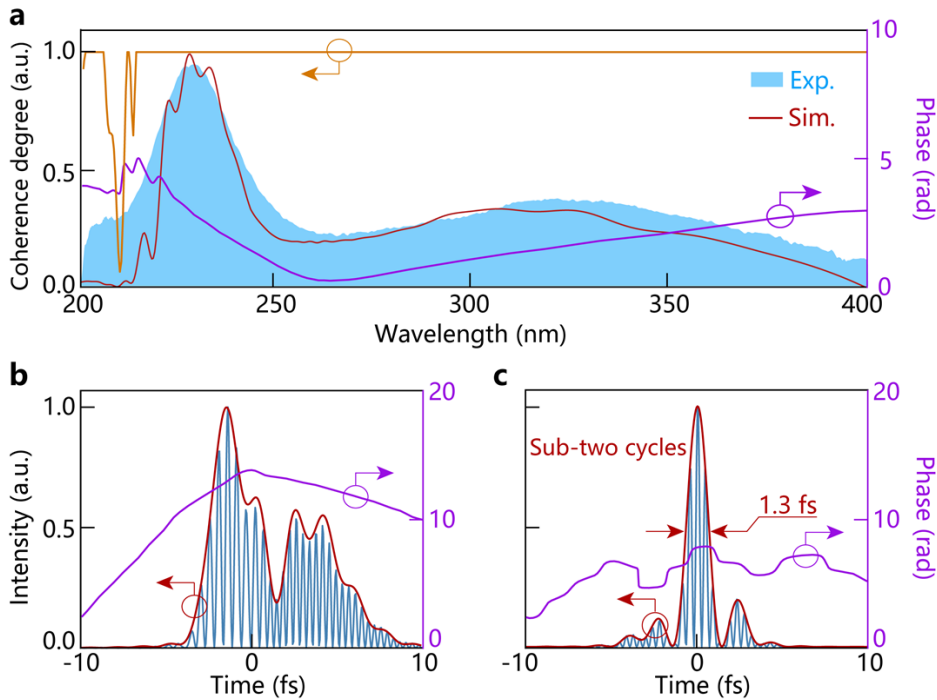

**Fig. 4 | The phase profile and coherence performance of the broadband UV DW. a**, Simulated phase profile (purple line) and complex degree of first-order coherence (yellow line) of the broadband UV DW. The measured (blue shadow) and simulated (red line) DW spectra are plotted on a linear scale, corresponding to the case at the pump pulse energy of 64 µJ in Fig. 3. **b,c**, Corresponding numerical simulations of the normalized temporal envelope (red lines), electrical field intensity

|E|$^2$ (blue lines) and phase profiles (purple lines). (**c**) is the compressed pulse through introducing -2 fs$^2$ group delay dispersion to the output DW pulse (**b**).

At a pump pulse energy of 64 µJ, the pulse energy of the broadband UV DW was experimentally measured to be 4.07 µJ (see Methods below), corresponding to an overall emission efficiency of ~6.4%. At this pump energy, the output DW exhibits good spectral stability. In the experiment, we measured the broadband DW spectrum for 1000 times with a time interval of 1.8 s and scanning time of 0.1 s, and the results of 30-minute recording are plotted in Fig. 5a. The relative root-mean-square energy fluctuation is estimated to be ~1.9% (see Methods below).

The good phase coherence and spectral stability of the broadband DW were further verified in the experiments of pulse chirp compensation and temporal width measurement. We constructed a self-diffraction (SD) FROG set-up and mounted it in a gas chamber, see Methods below and Section 1 of the Supplementary Information for some details of the diagnosis set-up. Two chirp mirrors were used in the chamber to compensate the pulse chirp accumulated over the optical path. When the gas pressure inside the chamber was fixed at 1 bar (atmospheric pressure), the gas type was air-helium mixture. Therefore, we could vary the ratio of helium gas in the mixture to adjust precisely the GDD value in the optical path, so as to minimize the residual pulse chirp at the measurement point. At a pulse energy of 56 µJ, the measured and retrieved FROG traces are illustrated in Figs. 5b-e, with a reasonable FROG retrieving error of ~0.7%. Note that the measured spectrum (plotted as black line) by the FROG set-up is much narrower than the original spectrum (plotted as blue shadow) measured at the output port of the capillary fibre (see Fig. 5e), which is due to the limited bandwidth of the two chirp mirrors (CM313 from Ultrafast Innovations, with a transmission window from 230 nm to 270 nm). The filtered UV spectrum under test has a FWHM bandwidth of ~40 nm, corresponding to a FTL pulse width of 3.2 fs (spectral range within 10-dB intensity). The retrieved FWHM pulse width is 3.7 fs (see Fig. 5d), quite close to this FTL value. Even though the measured UV spectrum is merely a small portion of the generated DW spectrum (see Fig. 5d), the good compensability of the UV output pulse (close to its FTL width) verifies, to some extent, the good coherence and smooth phase of the broadband DW.

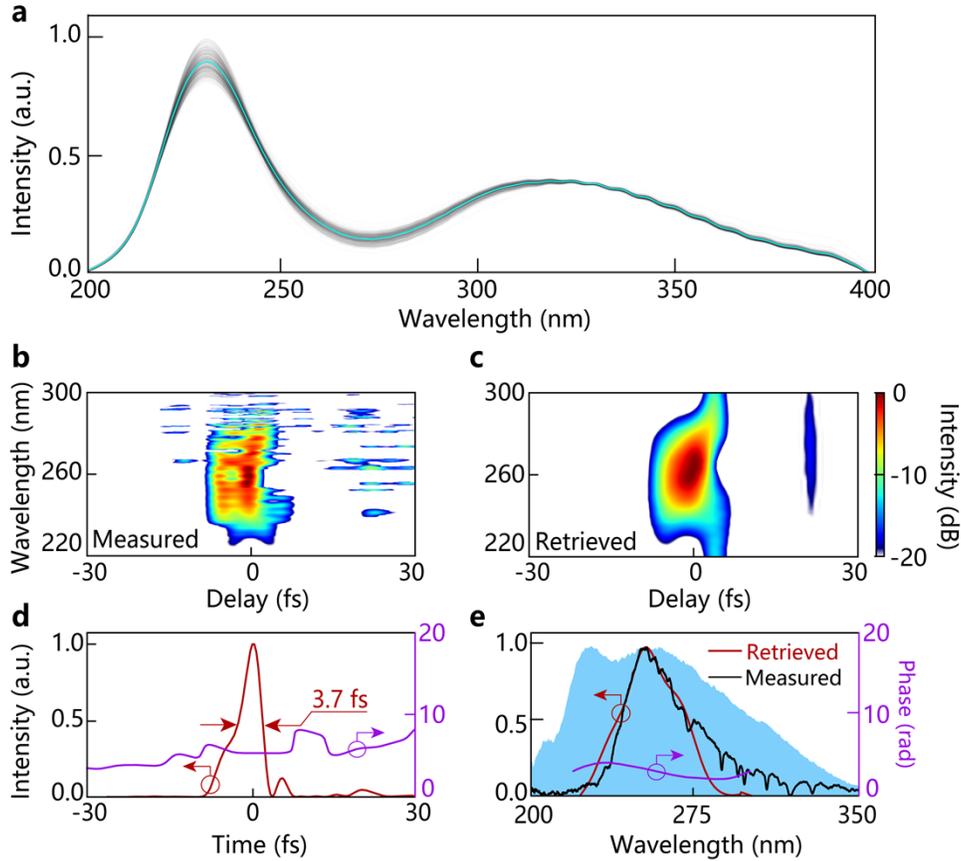

**Fig. 5 | Measured spectral stability and temporal characterization of the broadband UV DW. a**, Measured spectral stability of the broadband UV DW over 30-minute recording; gray shadow, 1000 measurement results with a time interval of 1.8 s and scanning time of 0.1 s; cyan line, the mean value of these results. The experimental parameters of pump pulse, HCF and gas are the same as those in Fig. 1c. **b**, Measured SD-FROG trace. **c**, Retrieved SD-FROG trace with a retrieving error of 0.7%. **d**, Retrieved temporal (red line) and phase (purple line) profiles. The DW pulse shows a temporal width of 3.7 fs. **e**, Retrieved spectral (red line) and phase (purple line) profiles. The experimental spectra are plotted as blue shadow and black line, measured before and after the chirped mirrors, respectively. In (**b-e**), the experimental parameters correspond to the case at the pump pulse energy of 56 µJ in Fig. 3c.

## Discussions and conclusion

We found that this spectral broadening phenomenon of DW, induced by soliton splitting dynamics, is generic, which can be observed over a wide range of system parameters. In the experiment, we varied the Ne-gas pressure in the capillary fibre from 3 bar to 7 bar, corresponding to a tuning of the zero GVD wavelength of the waveguide from 460 nm to 564 nm and a tuning of the phase-matching wavelength from 194 nm to 310 nm[6,50]. Note that the change of gas pressure can also vary Kerr nonlinearity of the waveguide as well as GVD value at the pump wavelength. Experimentally, we fixed the pump pulse width to 18 fs and adjusted accordingly the pump pulse energy, so as to maintain the soliton order to be ~7.4. Continuous central-wavelength tuning of the broadband DW emission was observed in the experiments (see Fig. 6), exhibiting similar tuning abilities as narrow-band DW emission experiments[6,10,11]. Simulation results show that this mechanism of DW broadening is also valid when using a different type of gas in the capillary. For example, when 18-fs, 210-µJ, 800-nm pump pulse (soliton order N = 7.4) was used, the 100-µm-inner-diameter, 35-cm-long capillary, filled with 3.2-bar He gas, could be used to extend the broadband DW emission to the vacuum UV regime, covering 140 – 240 nm wavelengths. See Section 5 of the Supplementary Information for some detailed results and discussions.

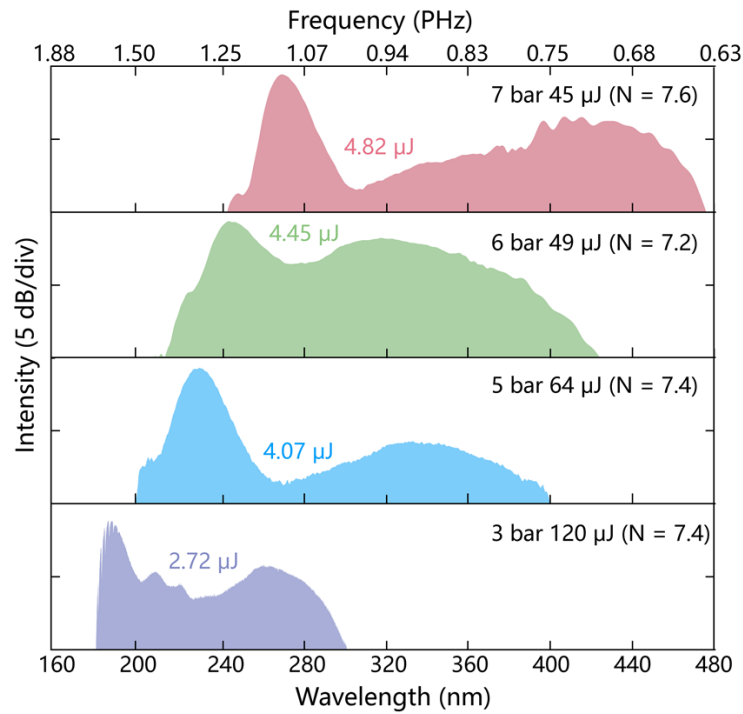

**Fig. 6 | Central-wavelength tuning of the broadband UV DW.** The broadband DW spectra with different emission wavelengths are measured at different Ne-gas pressure from 3 bar to 7 bar. The pump pulse used in the experiment has a temporal width of 18 fs and the HCF has a length of 42 cm and a core diameter of 100 μm. These results are also plotted on the coordinate axis of spectrum versus frequency, see Supplementary Fig. S13.

Even though interactions between Raman-shifted solitons and short-wavelength DWs have been demonstrated[46,47,63] (also known as "soliton trapping effect" observed during the supercontinuum generation process in solid-core photonic crystal fibre), the generation of octave-spanning UV DW with μJ-level pulse energy and good spectral coherence has never been explored before. The extreme spectral broadening due to strong cross-phase modulations, can be mainly attributed to two reasons. First, the secondary pulse generated due to soliton splitting dynamics, appears just behind the self-compressed pulse (with merely few-femtosecond temporal delay), and it has a relatively high peak power (see Figs. 2g,h) comparable to that of the self-compressed pulse, strongly enhancing the cross-phase-modulation effect. Second, the short delay time between the self-compressed and the secondary pulses ensures the fact that the UV DW, after being generated, immediately meets with the intense secondary pulse after only a short-length (~4 cm, see Figs. 2f,g) propagation in the capillary waveguide. Therefore, excessive temporal stretching of the DW pulse, due to waveguide dispersion, is efficiently suppressed, leading to a perfect temporal overlapping of the two pulses (see Fig. 2g). In contrast, the soliton trapping effect, demonstrated in supercontinuum generation[46,47,63], relies on the deceleration of Raman-frequency-shifted soliton which normally requires more propagation distance. This long-distance (tens of centimetres[46,47,63]) propagation results in large temporal stretching of the DW, significantly weakening the interaction strength between the two pulses.

Combining the advantages of μJ-level pulse energy, octave-wide spectral coverage and good spatial and spectral coherence, the ultra-broadband DW demonstrated here represents a major step forward to high-quality, single-cycle or even sub-cycle UV pulse generation, which is critical for many applications in ultrafast electron microscopy[64], femtosecond spectroscopy[65-67], advanced photolithography[68,69] and UV frequency comb[70,71]. The results also indicate that the introduction of two or even more driving pulses in the nonlinear optical system may present some new opportunities for observing unique dynamics and phenomena concerning nonlinearly-coupled propagation of multiple pulses[72,73], offering additional degrees of freedom for manipulating nonlinear optics experiments.

## Acknowledgements

This work was supported by the National Natural Science Foundation of China (62205353, 62275254, 61925507 and 12388102), the National Postdoctoral Program for Innovative Talents (BX2021328), the China Postdoctoral Science Foundation (2021M703325), the Strategic Priority Research Program of the Chinese Academy of Science (XDB0650000), the Shanghai Science and Technology Plan Project Funding (23JC1410100), the Shanghai Science and Technology Innovation Action Plan (21ZR1482700), National Key R&D Program of China (2022YFA1604401), Shanghai Science and Technology Committee Program (22DZ1100300, 22560780100 and 23560750200), Fuyang High-level Talent Group Project.

## Author contributions

Z.Y.H., M.P. and Y.X.L. conceived the work. T.D.C., Y.Y. and Z.Y.H. carried out the experiments. T.D.C., J.Y.P. and Z.Y.H. performed the numerical simulations. T.D.C., J.Y.P., Z.Y.H. and M.P. made the theoretical and experimental analysis and wrote the manuscript. All authors contributed to the discussion of the results and the editing of the manuscript.

## Competing interests

The authors declare no competing interests.

## Additional information

Supplementary information is available for this paper at xxxxxx.

## Methods

**Experimental set-up.** The broadband UV DW generation system consists of two stages, including HCF compression and broadband DW emission, driven by a commercial Ti: sapphire femtosecond laser, delivering 800 nm, 50 fs pulses at a repetition rate of 1 kHz. In the first stage, the pump pulses with an energy of 600 μJ were compressed using an HCF compression set-up, consisting of a 78-cm-long capillary fibre, with 250 μm inner diameter, filled with Ar gas, followed by a broadband attenuator and two pairs of chirped mirrors (103367, Layertec). The system can compress the input pulses from 50 fs to 18 fs (10 fs) at 130 mbar (330 mbar) Ar-gas pressure. The transmission efficiency of this system is ~40%, corresponding to a maximum output pulse energy of 240 μJ. In the second stage, another 42-cm-long Ne-filled capillary fibre with a core diameter of 100 μm was used to enable the broadband UV DW generation. See the Supplementary Fig. S1 for more details of the set-up.

**Pulse spectrum and energy measurements.** Measure the pulse spectrum in the wavelength range of 180 nm to 300 nm using a UV spectrometer (Maya2000-Pro, 155-310 nm, Ocean Insights), in the wavelength range of 200 nm to 400 nm using another UV spectrometer (Maya2000-Pro, 200-400 nm, Ocean Optics), in the wavelength range of 400 nm to 1000 nm using a spectrometer from UV to near-infrared region (Maya2000-Pro, 210-1100 nm, Ocean Optics). The last two spectrometers have been calibrated and can be used to measure the pulse spectrum within a wide spectral range from 200 nm to 1000 nm. By interpolating data within the overlapping area and combining the spectra measured by these two spectrometers, we finally obtained the full optical spectra at the HCF output port, as shown in Figs. 1b,c and Fig. 3a. It should be noted that in the spectral measurements, the laser pulse at the HCF output port was divided into two parts by a fused silica wedge. The transmitted pulse entered the integrating sphere connected to the fibre-coupled UV spectrometer (Maya2000-Pro, 200-400 nm, Ocean Optics), while the reflected pulse entered another integrating sphere connected to the fibre-coupled spectrometer (Maya2000-Pro, 210-1100 nm, Ocean Optics). Moreover, the broadband UV DW spectrum with a spectral range from 180 nm to 300 nm generated at Ne-gas pressure of 3 bar, as shown in Fig. 6, was measured using the UV spectrometer (Maya2000-Pro, 155-310 nm, Ocean Insights) placed in the $N_2$-filled chamber connected to the HCF output port.

The pulse energy at the input and output ends of the first and second stage HCF was measured using a thermal power meter (3A-P, Ophir). The energy measurements of the broadband UV DW pulses were achieved through the use of bandpass filters (Pelham Research Optical for 200-260 nm with 20 nm interval; FBH343-10, Thorlabs for 343 nm; FGUV11, Thorlabs for 275-375 nm), a photodiode (PD300-UV, Ophir), and a UV spectrometer (Maya2000-Pro, 200-400 nm, Ocean Optics). We performed the energy measurement of the broadband UV DW pulse generated at the input pulse energy of 64 μJ and Ne-gas pressure of 5 bar as an example. In the measurements, the bandpass filter was firstly used to filter out a portion of the spectrum of the broadband UV DW pulse, and the photodiode and UV spectrometer were used to measure the energy and spectral intensity of the filtered UV pulse. These results were used to calibrate the measured spectral energy density. Then the UV spectrometer was used to measure the spectral intensity of the broadband UV DW pulse before the bandpass filter. Finally, based on the calibrated spectral energy density and the spectral intensity before filtering, the energy of the broadband

UV DW pulse can be estimated to 4.07 µJ, see Fig. 3c. Using the same method, the energy of the UV DW pulses with different bandwidths and central wavelengths have also been measured, and the results are illustrated in Fig. 3c and Fig. 6.

**Beam profile measurements.** In Fig. 1d, near-field beam profiles of the broadband DW pulse were measured at different wavelengths, using several optical filters (Pelham Research Optical for 200-260 nm; FBH343-10, Thorlabs for 343 nm; FGUV11, Thorlabs for 275-375 nm) and a CCD camera (BGS-USB3-SP932U, Ophir-Spiricon).

**Numerical simulations.** The numerical simulations (see Figs. 1-4) were performed using both the GNLSE model and the single-mode UPPE model[52-54] (including photoionization term). The single-mode UPPE model can be expressed as

$$\frac{\partial \tilde{E}(z,\omega)}{\partial z} = i\left(\beta(\omega) - \frac{\omega}{v_g}\right)\tilde{E}(z,\omega) - \frac{\alpha(\omega)}{2}\tilde{E}(z,\omega) + i\frac{\omega^2 \tilde{P}_{NL}(z,\omega)}{2c^2\varepsilon_0\beta(\omega)} \quad (2)$$

where $\tilde{E}(z,\omega)$ is the electric filed in frequency domain, $\beta(\omega)$ the propagation constant, $v_g$ the group velocity, $\alpha(\omega)$ the linear loss of the fibre, $c$ the light speed in vacuum, $\varepsilon_0$ the vacuum permittivity, and $\tilde{P}_{NL}(z,\omega)$ the nonlinear response in the frequency domain. In the simulations, we used the measured pulses as input and considered the dispersion introduced by a 0.5-mm-thick fused silica window. The numerical results simulated by the UPPE model show excellent agreements with those simulated by the GNLSE model. See Section 2 of the Supplementary Information for more details.

**Calculation of the complex degree of first-order coherence.** In Fig. 4a, 50 numerical simulations with different quantum noise (the simulation parameter are the same as those in Fig. 1c), performed using the UPPE model, can be used to calculate the complex degree of first-order coherence of the broadband UV DW, which can be given as[62]

$$\left|g_{mn}^{(1)}(\omega)\right| = \left|\frac{\langle \tilde{E}_m^*(\omega)\tilde{E}_n(\omega)\rangle}{\langle |\tilde{E}_m(\omega)|^2\rangle}\right| \quad (3)$$

where $\tilde{E}_m(\omega)$ and $\tilde{E}_n(\omega)$ are two different electric fields in the frequency domain and $\tilde{E}_m^*(\omega)$ is the complex conjugate of electric field $\tilde{E}_m(\omega)$. The angle brackets denote ensemble averaging over the independent simulation results.

**Estimation of the relative root-mean-square energy fluctuation.** In the experiment, the broadband UV DW spectrum was measured for 1000 times with a time interval of 1.8 s and scanning time of 0.1 s, as illustrated in Fig. 5a. The root-mean-square energy fluctuation can be calculated using

$$\delta_{rms} = \sqrt{\sum_{n=1}^{1000}\frac{(S_n - \bar{S})^2}{1000}} \quad (4)$$

where $S_n$ is the spectral area (equivalent to pulse energy) of the $n^{th}$ broadband UV DW pulse, and $\bar{S}$ is the average value of these 1000 measured spectra. The relative root-mean-square energy fluctuation of the broadband UV DW pulse can be estimated through $\delta_{rms}/\bar{S}$.

**Pulse characterization measurements**. Temporal characterization of 10 fs and 18 fs pump pulses output from the HCF compression stage was based on a home-built second-harmonic generation (SHG) FROG set-up with all reflective optics. In this set-up, the sum-frequency signal was generated in a 10-µm-thick beta barium borate (BBO) crystal cut for type I phase-matching. Pulse characterization measurement of broadband UV DW pulse at 250 nm was carried out using a home-built SD-FROG set-up. While the optical layout of the SD-FROG is similar to that of the SHG-FROG, a 50-µm-thick fused silica was used as the nonlinear medium and several aluminum mirrors were used to provide high reflections for the UV pulses. The SD-FROG set-up was placed in a gas chamber filled with air-helium mixture. Two chirp mirrors were used in the chamber to compensate the pulse chirp accumulated over the optical path. In addition, the GDD value in the optical path was precisely adjusted by varying the ratio of helium gas in the mixture, thereby minimizing the residual pulse chirp and therefore shortening UV DW pulse width at the measurement point, see Section 1 of the Supplementary Information for more details of the diagnosis set-up. The results of ultrafast pulse characterizations using the SHG-FROG and SD-FROG set-ups are illustrated in Section 4 of the Supplementary Information.

## Data availability

The data that support the plots within the paper and other findings of the study are available from the corresponding authors upon reasonable request.

## Code availability

The code used in this paper is available from the corresponding author upon reasonable request.

# Supplementary Information for

## "Octave-wide broadening of ultraviolet dispersive wave driven by soliton-splitting dynamics"

Tiandao Chen[1,4,5,†], Jinyu Pan[1,3,4,†], Zhiyuan Huang[1,2,\*], Yue Yu[1,3,4], Donghan Liu[1,2,4], Xinshuo Chang[1,3,4], Zhengzheng Liu[1], Wenbin He[2], Xin Jiang[2], Meng Pang[1,2,3,\*], Yuxin Leng[1,3,\*] and Ruxin Li[1,5]

[1]State Key Laboratory of High Field Laser Physics and CAS Center for Excellence in Ultra-intense Laser Science, Shanghai Institute of Optics and Fine Mechanics (SIOM), Chinese Academy of Sciences (CAS), Shanghai 201800, China

[2]Russell Centre for Advanced Lightwave Science, Shanghai Institute of Optics and Fine Mechanics and Hangzhou Institute of Optics and Fine Mechanics, Hangzhou, 311421, China

[3]Hangzhou Institute for Advanced Study, University of Chinese Academy of Sciences, Hangzhou 310024, China

[4]Center of Materials Science and Optoelectronics Engineering, University of Chinese Academy of Sciences, Beijing 100049, China

[5]Zhangjiang Laboratory, Shanghai 201210, China

\*Corresponding author: huangzhiyuan@siom.ac.cn; pangmeng@siom.ac.cn; lengyuxin@siom.ac.cn

†These authors contribute equally to the work

# Section 1: Experimental set-up

As shown in Supplementary Fig. S1, the experimental set-up for the broadband ultraviolet (UV) dispersive wave (DW) generation system includes two stages: hollow capillary fibre (HCF) compression and broadband UV DW emission. This system is driven by a commercial Ti: sapphire femtosecond laser (Solstice Ace 80L-35F-1K-HP-T, Spectra-Physics), producing 800 nm, 600 µJ, 50 fs pulses with a repetition rate of 1 kHz. The whole experimental system is installed on an optical table, and the physical size of this system is 2 m × 1 m. In the first stage, the 800 nm pulses from the driving laser were first launched in the Ar-filled HCF using a coated plano-convex lens with a focal length of 75 cm. The input pulse energy was controlled by a broadband attenuator base on a half-wave plate (HWP) and a wire grid polarizer (WGP). The HCF has a core diameter of 250 µm and a length of 78 cm. The input and output ends of the HCF were sealed with 0.5-mm-thick coated fused silica (FS) windows. The laser pulses experienced a spectral broadening in the Ar-filled HCF due to self-phase modulation. The pulse energy at the output port of the HCF was measured to be ~420 µJ, corresponding to a transmission efficiency of ~70%. The output pulses were collimated by a coated concave mirror with a focal length of 50 cm, and the pulse energy was controlled by another broadband attenuator, as shown in Supplementary Fig. S1. Two pairs of chirped mirrors (103367, Layertec) were used to compensate for the output pulses. When the HCF was filled with 130 mbar or 330 mbar Ar gas, the laser pulses can be compressed from 50 fs to 18 fs or 10 fs, respectively, measured using a home-built second-harmonic generation frequency-resolved optical gating (SHG-FROG) set-up. The efficiency of the HCF compression stage is ~40%, corresponding to a maximum pulse energy of 240 µJ.

In the second stage, the compressed pulses from the first stage were coupled in another Ne-filled HCF with a core diameter of 100 µm and a length of 42 cm, using a parabolic mirror with a focal length of 20.3 cm. The input and output ports of the second HCF were sealed with a 0.5-mm-thick coated FS window and a 0.2-mm-thick uncoated magnesium fluoride ($MgF_2$) window, respectively. The entire HCF configuration can operate stably at gas pressure as high as 7 bar. The efficiency of the broadband UV DW emission stage is ~31%, including ~75% coupling efficiency and ~41% transmission efficiency. The broadband UV DW pulses generated at the output of the second HCF will be diagnosed, including time-domain, spectral, energy and beam profile measurements.

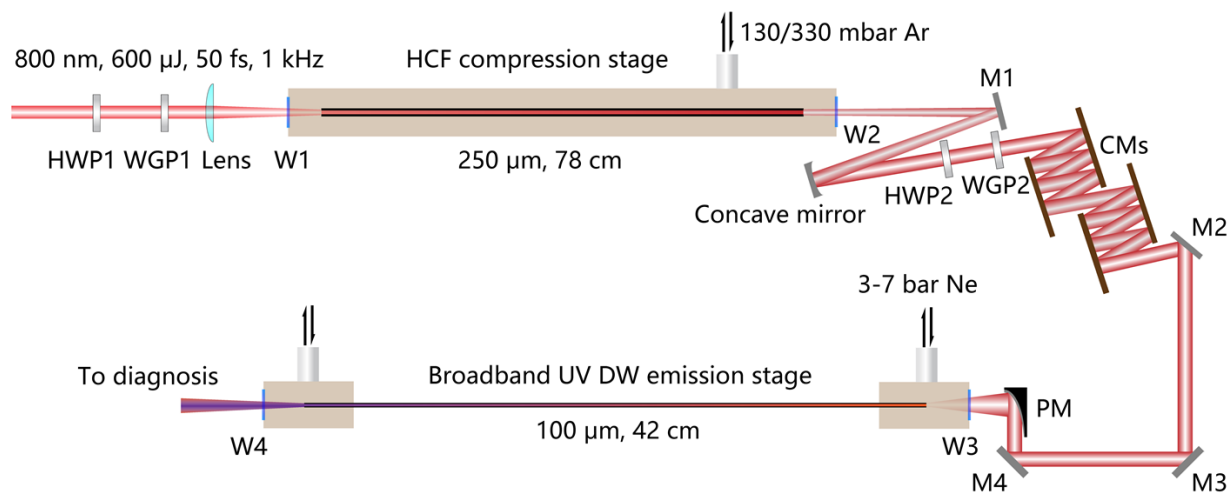

**Supplementary Fig. S1. Experimental set-up.** HWP1-HWP2, half-wave plate; WGP1-WGP2, wire grid polarizer; W1-W4, the first three (W1-W3) are fused silica (FS) windows, and the last one (W4) is magnesium fluoride ($MgF_2$) widow; HCF,

hollow capillary fibre; M1-M4, silver mirrors; CMs, chirped mirrors; PM, parabolic mirror; UV DW, ultraviolet dispersive wave.

The time-domain measurements of the broadband UV DW pulses were performed using a home-built self-diffraction frequency-resolved optical gating (SD-FROG) set-up mounted in a gas chamber, as shown in Supplementary Fig. S2. The input and output ports of the gas chamber were sealed with a 0.1-mm-thick uncoated FS window and a 0.5-mm-thick uncoated $MgF_2$ window, respectively. The gas pressure inside the chamber was fixed at 1 bar (atmospheric pressure), the gas type was Air-He mixture. In the gas chamber, the output pulses from the second HCF first passed through two chirped mirrors (CM313, Ultrafast Innovations) with high reflectivity over the spectral range of 230 to 270 nm, which were used to filter out the residual pump pulse and compensate for the broadband UV DW pulse. The pulses were then divided into two parts by the first D-shaped mirror (DM1). These two pulses were focused into a 50-μm-thick FS plate using a concave mirror with a focal length of 10 cm, which excited a transient grating and generated a self-diffraction (SD) signal. The SD signal was collimated by a parabolic mirror with a focal length of 15.3 cm and finally focused into a fibre-coupled UV spectrometer (Maya2000-Pro, 200-400 nm, Ocean Optics) using an uncoated plano-convex calcium fluoride ($CaF_2$) lens with a focal length of 5 cm.

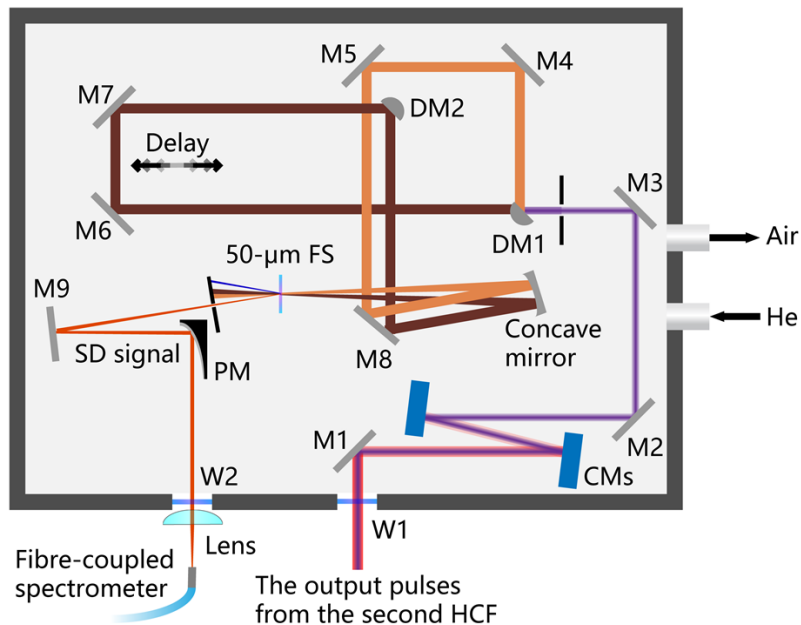

**Supplementary Fig. S2. Layout of the SD-FROG device.** W1-W2, W1 is fused silica (FS) window and W2 is magnesium fluoride ($MgF_2$) window; M1-M9, UV-enhanced aluminium mirrors; CMs, chirped mirrors; DM1-DM2, D-shaped mirrors. PM, parabolic mirror; SD, self-diffraction.

Before generating the SD signal, broadband UV DW pulses accumulated dispersion as they passed through the optical path. Supplementary Table S1 provides the dispersion sources through which UV pulses passed and the estimated dispersion amount. The broadband UV DW was generated very close to the output port of the second HCF, which is approximately 15 cm away from the 0.2-mm-thick $MgF_2$ window mounted at the output end of the HCF. As shown in Supplementary Table S1, the group delay dispersion (GDD) provided by 15-cm-long Ne gas at 5 bar and 0.2-mm-thick $MgF_2$ window are 8 $fs^2$ and 21 $fs^2$, respectively. The gas chamber was located 10 cm away from of the HCF output port, resulting in a GDD value of 12 $fs^2$ caused by 10-cm-long Air. When UV pulses entered the gas chamber, the 0.1-mm-thick FS window installed at the input port of the gas chamber introduces 23 $fs^2$ GDD. Two chirped mirrors in the gas chamber introduce -100 $fs^2$ GDD. Moreover, in the FROG set-up, the length of free space filled with a mixture of Air and He gas was measured to be ~1.5 m, leading to a GDD range of 6 $fs^2$ (100% He gas) to 177 $fs^2$ (100% Air). Therefore, the total GDD was estimated to be -30 to 141 $fs^2$, as shown in Supplementary Table S1. In the experiment, we could vary the ratio of He gas in the mixture to adjust precisely the GDD value in the optical path, thereby minimizing the residual pulse chirp at the measurement point and obtaining the shortest pulse width.

| Dispersion source | GDD ($fs^2$) |
|---|---|
| 15-cm-long Ne gas at 5 bar | 8[1] |
| 0.2-mm-thick $MgF_2$ window | 21[1] |
| 10-cm-long Air | 12[1] |
| 0.1-mm-thick FS window | 23[1] |
| 2 chirped mirrors | -100 |
| 1.5-m-long Air-He mixture at 1 bar | From 6[2] (100% He) to 177[1] (100% Air) |
| Total | From -30 to 141 |

**Supplementary Table S1. Dispersion values of different elements at 250 nm.**

## Section 2: Simulation models

The experimental results of the broadband UV DW emission process were reproduced using the generalized nonlinear Schrödinger equation (GNLSE) and the single-mode unidirectional pulse propagation equation (UPPE). In the simulations, we used the measured pulses as input, taking into account the dispersion introduced by a 0.5-mm-thick FS window mounted at the input port of the HCF. We found that the numerical results of soliton self-compression and broadband UV DW emission processes in the capillary fibre, simulated by these two models, exhibit excellent agreement with each other, as shown in Supplementary Fig. S3. The main difference between these two models is that the GNLSE model cannot describe the third-harmonic generation (THG), see Supplementary Fig. S3.

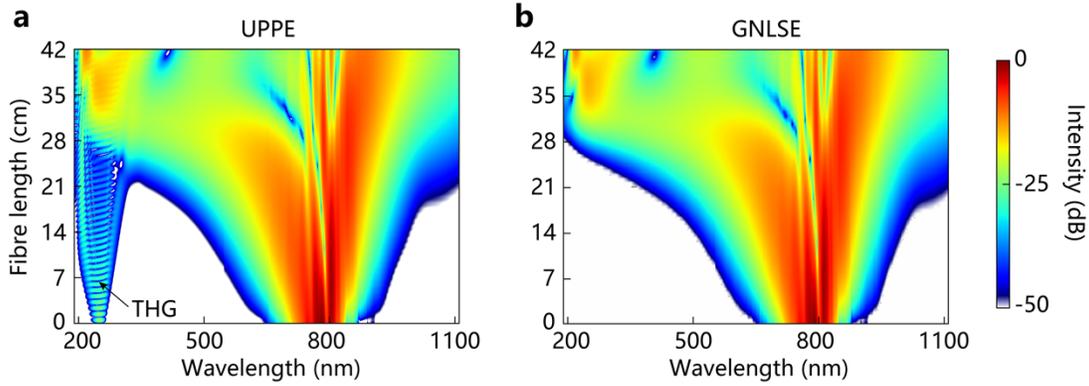

**Supplementary Fig. S3. Spectral evolution of the pulse propagating in the HCF.** (a) Simulated spectral evolution of the 18 fs, 64 μJ pulse (N = 7.4) using the UPPE model, with the same parameters as Fig. 2e in the main text. The third-harmonic generation (THG) is marked with the black arrow. (b) Simulated spectral evolution of the 18 fs, 58 μJ pulse (N = 7.0) using the GNLSE model, without considering gas ionization effect. The input pulse energy in (b) is slightly lower than that in (a), in order to better match the numerical simulations, all other simulation parameters are the same.

In addition, we also used the UPPE model to investigate the influence of the gas ionization effect on the broadband UV DW emission process. In the Eq. (2) of the main text, the nonlinear response can be expressed as[3]

$$\tilde{P}_{NL}(z,\omega) = F\left[\varepsilon_0 \chi^{(3)} E(z,t)^3 + P_{ion}(z,t)\right] \tag{S1}$$

where $F$ represents the Fourier transform, $\chi^{(3)}$ is the third-order nonlinear susceptibility that is relative to the Kerr effect, and $P_{ion}(z,t)$ is the ionization-induced plasma polarization, which can be given by[3]

$$P_{ion}(z,t) = \int_{-\infty}^{t} \frac{\partial \rho(z,t')}{\partial t'} \frac{U_i}{E(z,t')} dt' + \frac{e^2}{m_e} \int\int_{-\infty}^{t} \rho(z,t') E(z,t') dt' dt' \tag{S2}$$

where $\rho$ is the plasma density, $U_i$ is the ionization potential of the filled gas, $e$ is the electron charge and $m_e$ is the electron mass. The plasma density $\rho$ inside the HCF core, due to gas ionization, was calculated using the Perelomov-Popov-Terent'ev model[4], modified with the Ammosov-Delone-Krainov coefficients[5].

As shown in Supplementary Figs. S4a,b, we plotted the spectral evolution of the pulse propagating in the HCF with and without considering gas ionization effect. We can see that the gas ionization process has a negligible impact on the spectral broadening of the pulse and the broadband UV DW emission process.

In addition, when the pulse propagated in the HCF, the ionization fraction of Ne gas is always below the level of $10^{-5}$, which further indicates that gas ionization effect can be neglected in the generation process of broadband UV DW, see Supplementary Fig. S4c.

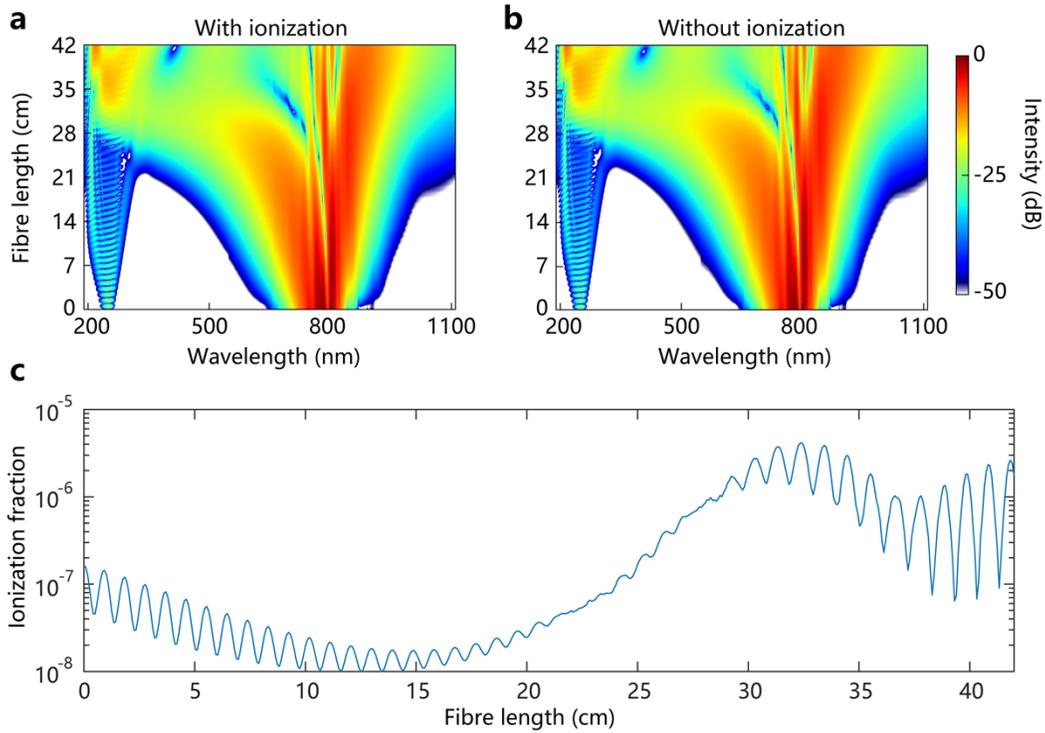

**Supplementary Fig. S4. Influence of gas ionization.** Simulated spectral evolution of the 18 fs, 64 µJ pulse (N = 7.4) using the UPPE model with ionization turned on (a) and with ionization turned off (b), with the same parameters as Fig. 2e in the main text. (c) Ionization fraction as a function of fibre length for (a).

# Section 3: Asymmetry in nonlinear pulse propagation

As mentioned in the main text, the higher-order dispersion, self-steepening effect, wavelength-dependent capillary loss and photoionization effect could be regarded as some perturbations to propagation processes of canonical solitons described by the Kerr nonlinearity and 2nd-order dispersion terms. Near the maximum soliton self-compression point, these perturbations could cause the highly-asymmetric soliton splitting phenomenon. In Section 2, the numerical simulation results indicate that the gas ionization effect can be neglected without inducing any influence on the broadband UV DW emission. In this Section, we numerically investigated the influences of higher-order dispersion, self-steepening effect, wavelength-dependent capillary loss and pump pulse shape on broadband UV DW emission using the GNLES model. The GNLSE model has been proven to be equally effective as the UPPE model in simulating the dynamic processes of soliton self-compression and broadband UV DW emission in the gas-filled HCF, see Supplementary Fig. S3 in Section 2. Neglecting the gas ionization effect, the GNLES can be rewritten as

$$\frac{\partial A}{\partial z} + \frac{i\beta_2}{2}\frac{\partial^2 A}{\partial T^2} - i\gamma|A|^2 A = \sum_{n=3}^{\infty} i^{n+1} \frac{\beta_n}{n!}\frac{\partial^n A}{\partial T^n} - \frac{\gamma}{\omega_0}\frac{\partial}{\partial T}(|A|^2 A) - \frac{\alpha}{2}A \qquad (S3)$$

As shown in Supplementary Fig. S5, we performed numerical simulations using a perfect Gaussian-shaped pump pulse with the same energy and similar temporal width as shown in Fig. 2e of the main text. The simulated broadband UV DW generation process is very similar as that shown in Figs. 2e-h of the main text, see Supplementary Movie 3 for more detailed information. Therefore, the slight change of the pump pulse shape has little impact on its nonlinear propagation dynamics.

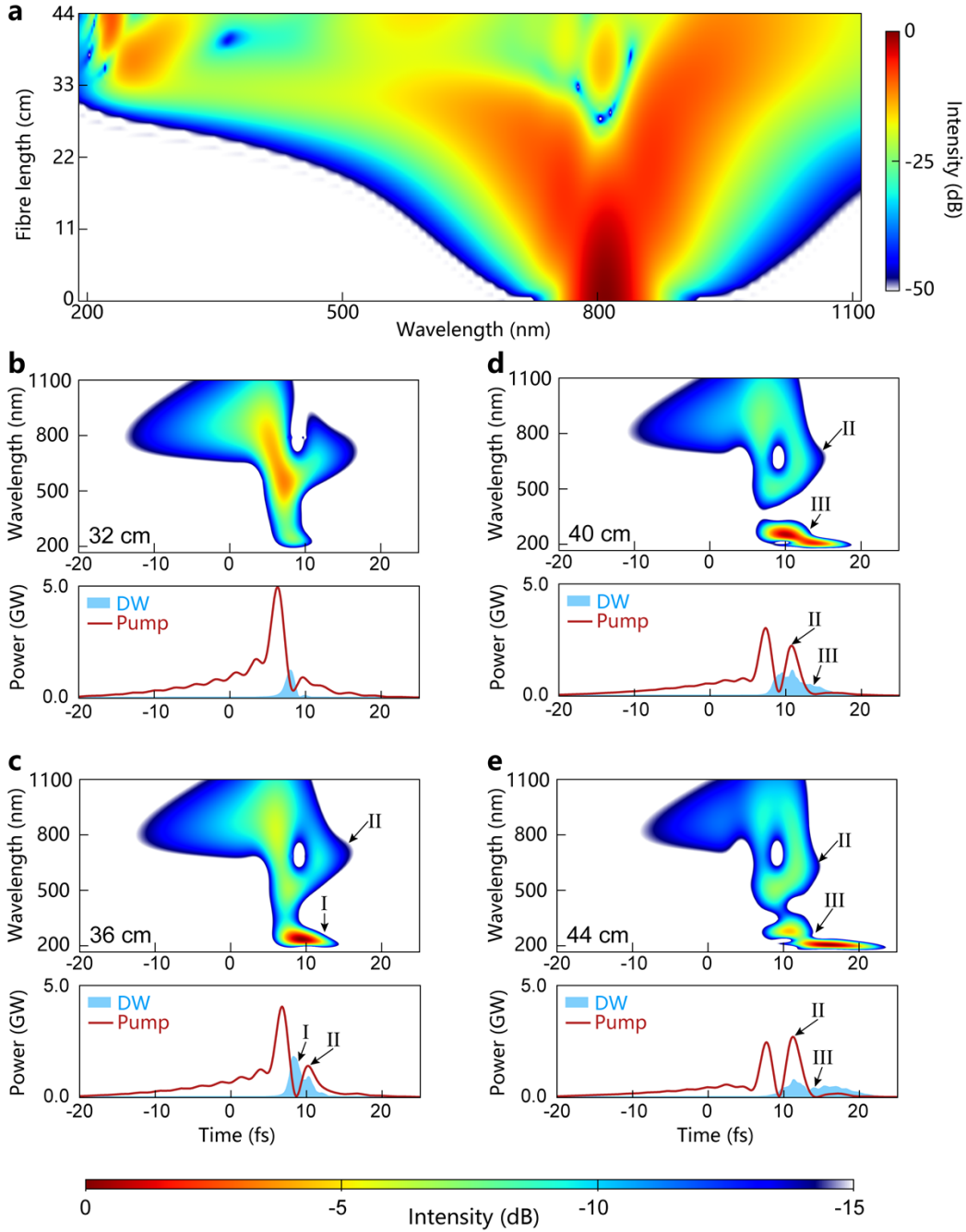

**Supplementary Fig. S5. Spectral evolution and spectrogram analysis of Gaussian-shaped pulse propagating in the HCF.** (a) Simulated spectral evolution of the 19.5 fs, 64 μJ Gaussian-shaped pulse (N = 7.6) as a function of fibre length. Except for the fibre length set to 44 cm, all other simulation parameters are the same as Fig. 2e in the main text. (b-e) Spectrogram analysis of the pulse at the indicated positions in the HCF. **I** – Narrow-band DW emission. **II** – asymmetric soliton splitting phenomenon, **III** – spectral broadening of the UV DW through cross-phase modulation. Through using the ideal filters with transmission windows covering 400 to 4000 nm and 100 to 400 nm, the temporal profiles of the pump pulse and DW were obtained, respectively, which were plotted as red lines and blue shadows.

In the Eq. (S3), we set the capillary loss $\alpha$ as a constant $\alpha_0$ independent of the pulse wavelength. The capillary loss was calculated to be 9.624 dB/m at the pump wavelength of 800 nm for the 100-μm-inner-diameter HCF, which was used as the constant loss $\alpha_0$ for all wavelengths in numerical simulations, all other simulation parameters are the same as Supplementary Fig. S5. As shown in Supplementary Fig. S6, we can see that the simulated broadband UV DW generation process is very similar as that shown in Supplementary Fig. S5, see Supplementary Movie 4 for more detailed information. The main difference is that the intensity of the UV DW pulse is slightly lower than that in Supplementary Fig. S5. This is because the capillary loss is proportional to the square of the pulse wavelength[6] (based on this relation $\alpha \sim \lambda^2$). In numerical simulations, we used a capillary loss of 9.624 dB/m for all wavelengths, which is higher than the true loss in the short-wavelength spectral region. These results indicate that the perturbation induced by the capillary loss is not the cause of the highly-asymmetric soliton splitting phenomenon.

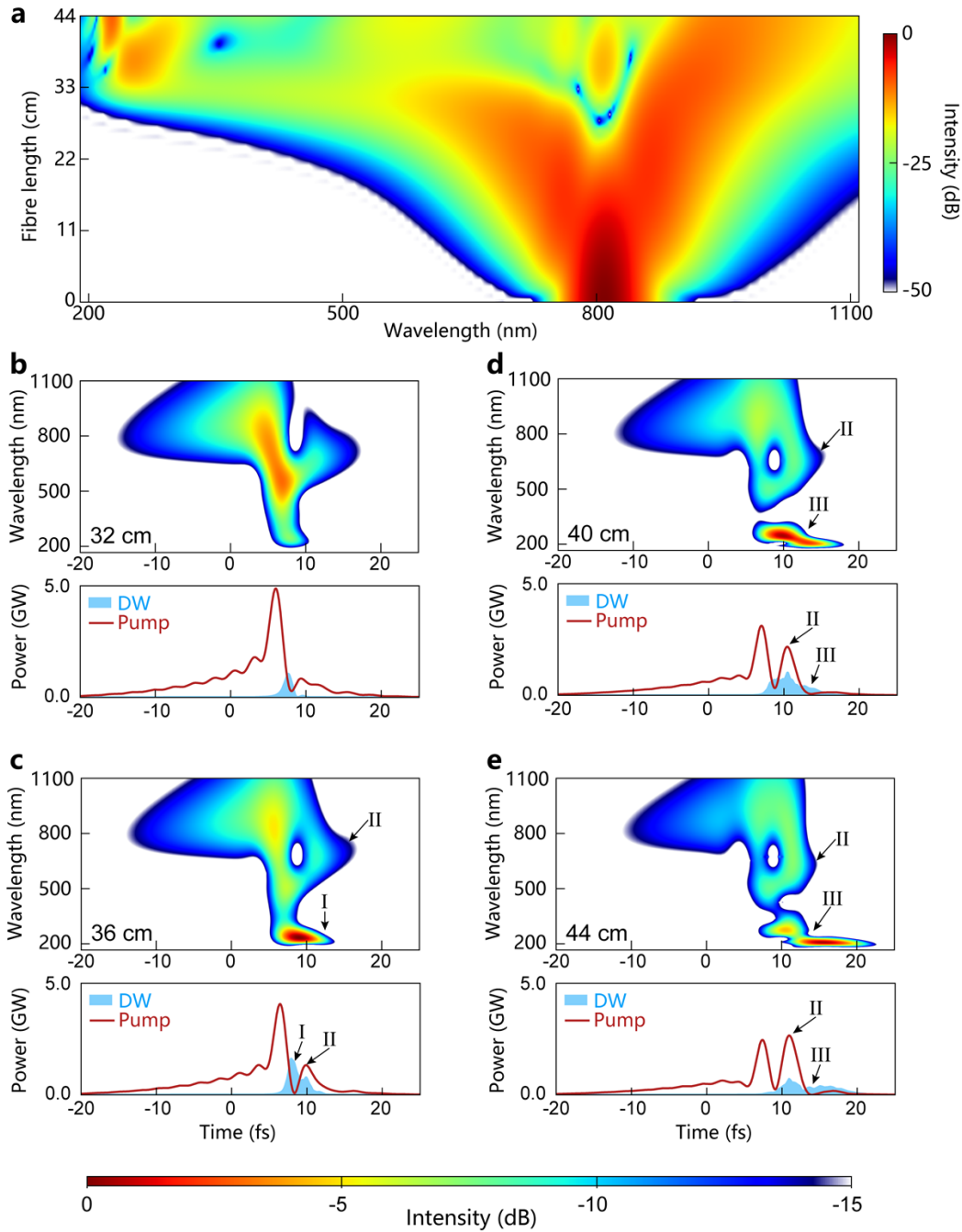

**Supplementary Fig. S6. Spectral evolution and spectrogram analysis of Gaussian-shaped pulse propagating in the HCF.** The capillary loss is set to a wavelength-independent constant, all other simulation parameters are the same as Supplementary Fig. S5.

When the self-steepening effect is not included in the simulation, the GNLSE model can be rewritten as

$$\frac{\partial A}{\partial z} + \frac{i\beta_2}{2}\frac{\partial^2 A}{\partial T^2} - i\gamma |A|^2 A = \sum_{n=3}^{\infty} i^{n+1} \frac{\beta_n}{n!} \frac{\partial^n A}{\partial T^n} - \frac{\alpha}{2} A \tag{S4}$$

We performed numerical simulations using the GNLSE model in Eq. (S4), the corresponding results are shown in Supplementary Fig. S7. Even if the self-steepening effect is turned off, we can still observe the broadband UV DW emission. However, compared with Supplementary Fig. S5, the broadband UV DW pulses generated in Supplementary Fig. S7 significantly decrease in DW emission efficiency. See Supplementary Movie 5 for more detailed information. Therefore, in the nonlinear dynamics, the self-steepening-induced blue shift of the pump pulse, accumulated over the soliton self-compression process, enhances the radiation efficiency of the short-wavelength DW.

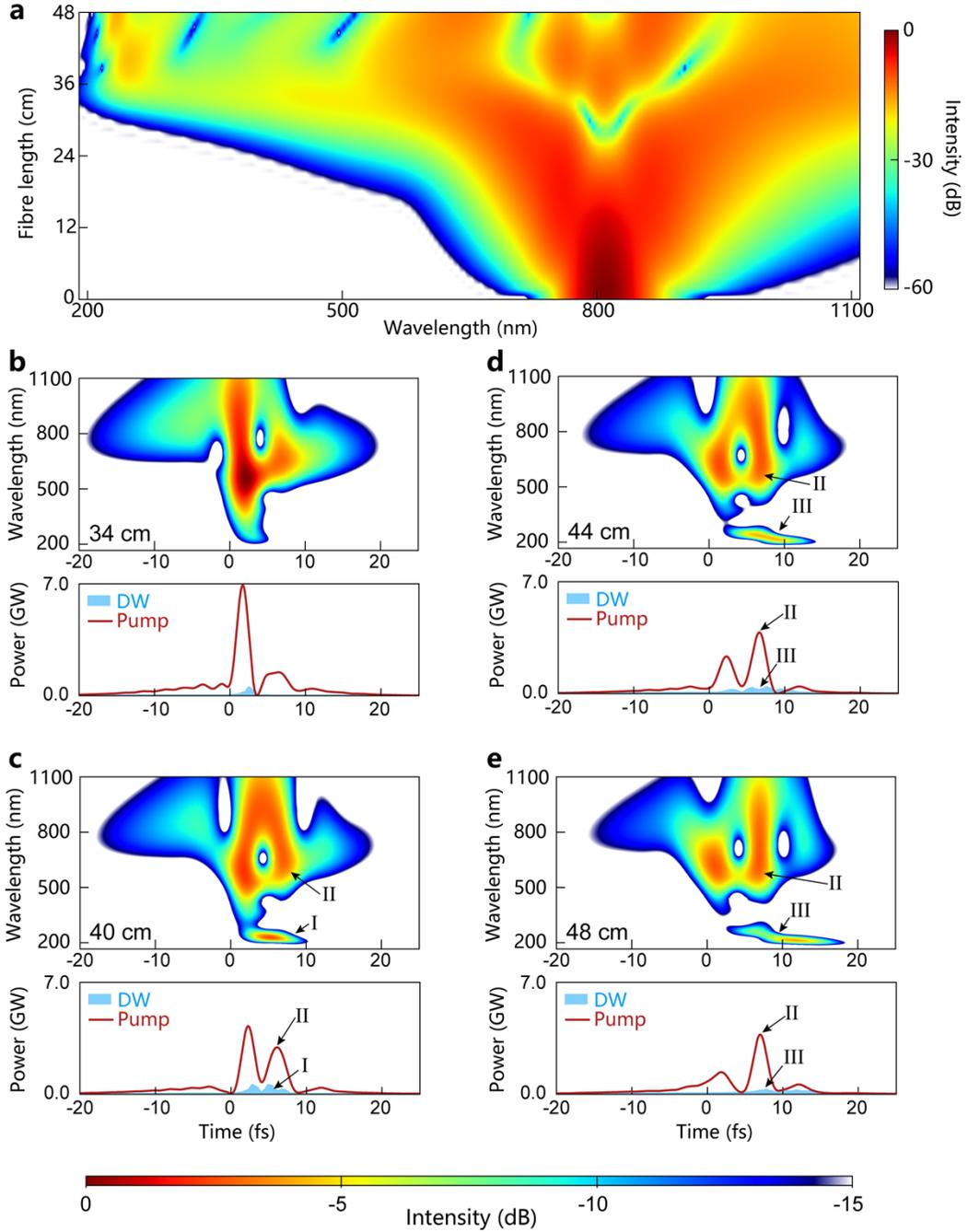

**Supplementary Fig. S7. Spectral evolution and spectrogram analysis of Gaussian-shaped pulse propagating in the HCF, with self-steepening effect turned off.** Except for the fibre length set to 48 cm, all other simulation parameters are the same as Supplementary Fig. S5.

In numerical simulations, the higher-order dispersion term is not included in the GNLSE model, Eq. (S3) can be rephrased as

$$\frac{\partial A}{\partial z} + \frac{i\beta_2}{2}\frac{\partial^2 A}{\partial T^2} - i\gamma |A|^2 A = -\frac{\gamma}{\omega_0}\frac{\partial}{\partial T}(|A|^2 A) - \frac{\alpha}{2}A \tag{S5}$$

As shown in Supplementary Fig. S8, there is no DW emission throughout the entire nonlinear pulse propagation process in HCF, see Supplementary Movie 6 for more detailed information. This can be explained as: the phased-matching DW emission is the result of the combined effects of self-compressed soliton and higher-order dispersion. The higher-order dispersion term plays a crucial role in the generation process of broadband UV DW emission.

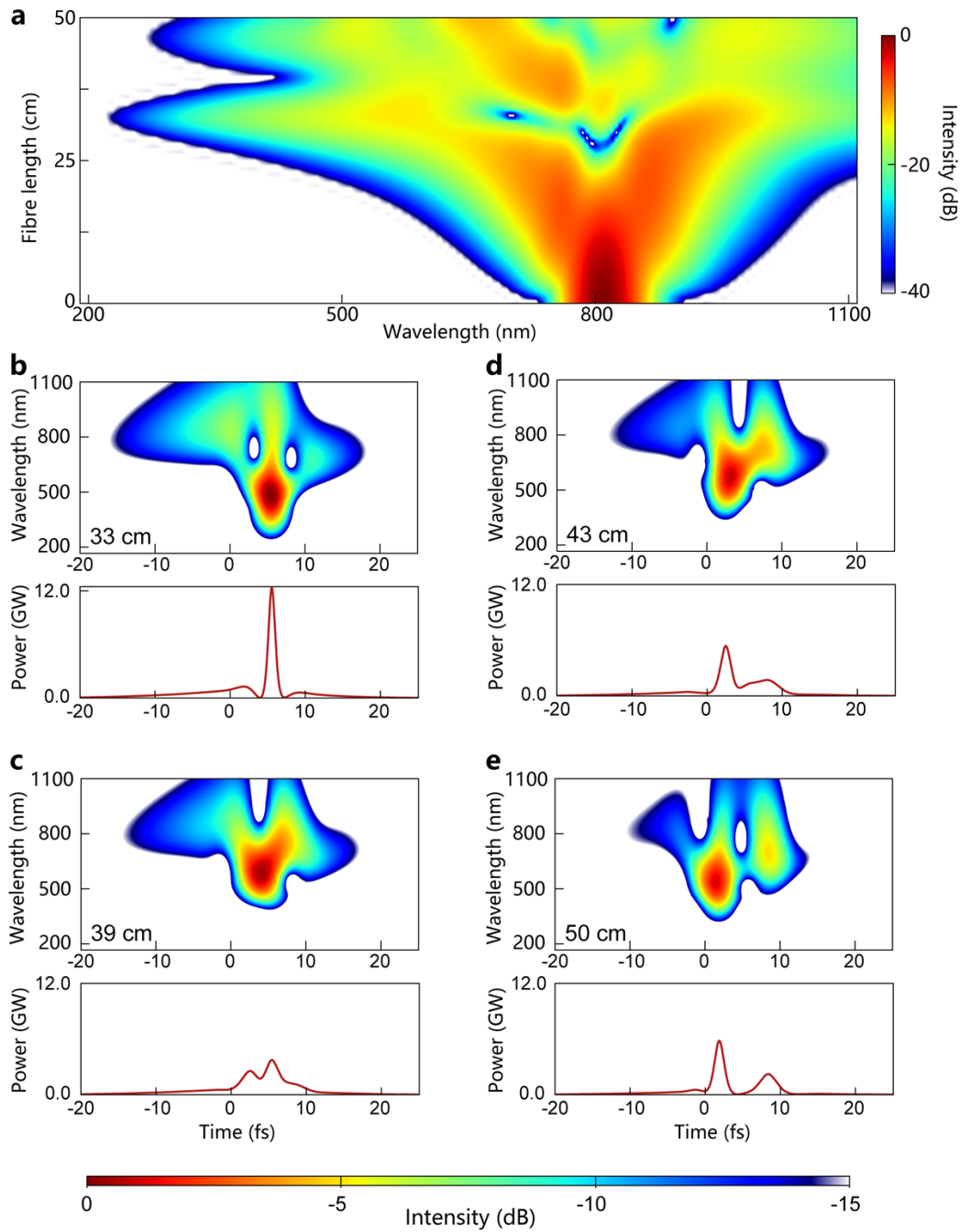

**Supplementary Fig. S8. Spectral evolution and spectrogram analysis of Gaussian-shaped pulse propagating in the HCF, with higher-order dispersion turned off.** Except for the fibre length set to 50 cm, all other simulation parameters are the same as Supplementary Fig. S5.

When the self-steepening effect and higher-order dispersion are not included in the simulation, and the capillary loss is independent on the pulse wavelength, the GNLSE model can be rewritten as

$$\frac{\partial A}{\partial z} + \frac{i\beta_2}{2}\frac{\partial^2 A}{\partial T^2} - i\gamma |A|^2 A = -\frac{\alpha_0}{2}A \tag{S6}$$

In Supplementary Fig. S9, we can see that there is no DW emission during the process of soliton self-compression, and the pulses exhibit good symmetry in both temporal and spectral distribution, see Supplementary Movie 7 for more detailed information. In summary, the numerical results in Section 3 indicate that the presence of higher-order dispersion directly determines the generation of DW pulse, and the self-steepening effect can significantly improve the emission efficiency of DW pulse. In particular, near the maximum soliton self-compression point, the combined influences of self-steepening effect and higher-order dispersion can result in highly-asymmetric soliton splitting phenomenon and high-efficiency broadband UV DW generation.

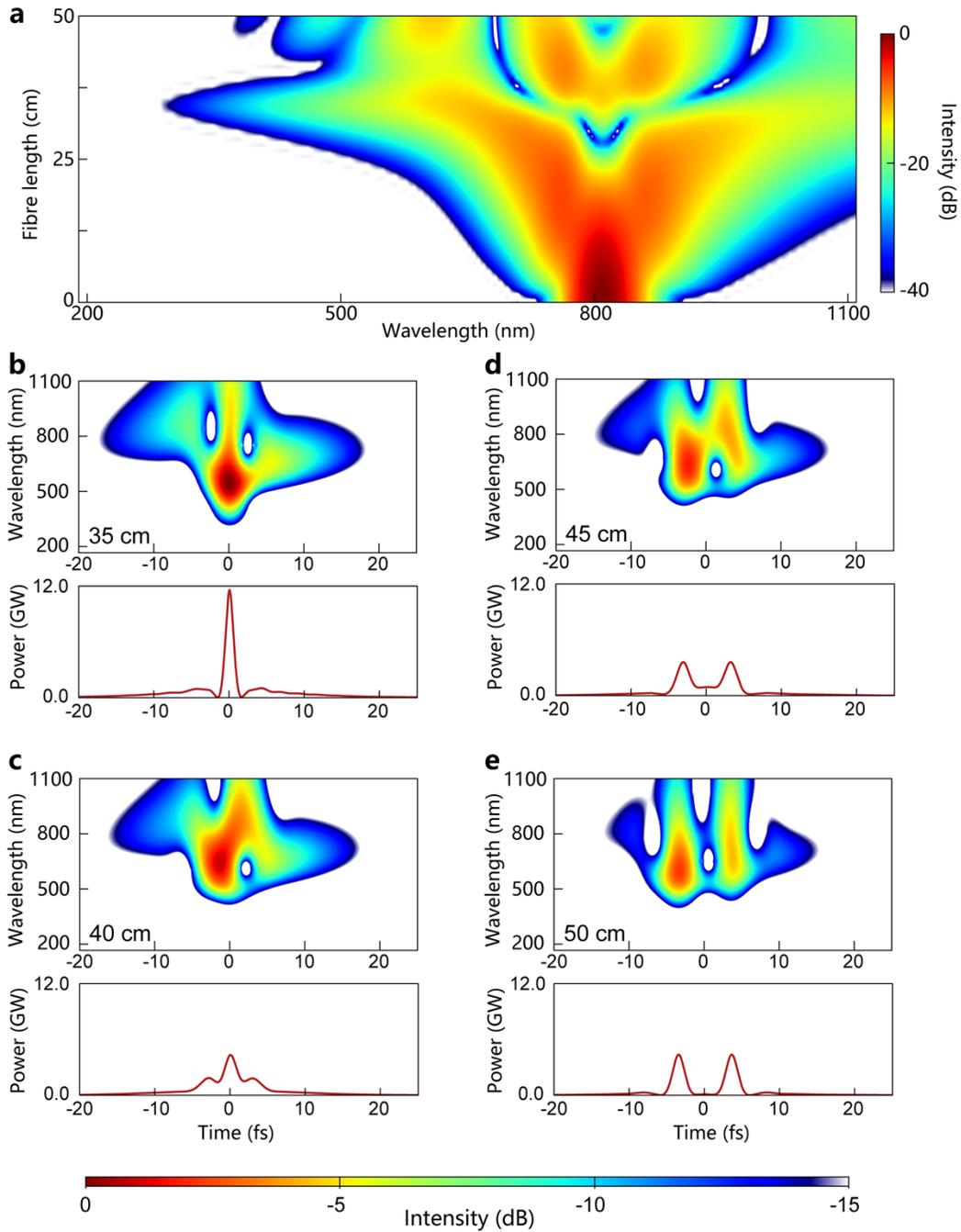

**Supplementary Fig. S9. Spectral evolution and spectrogram analysis of Gaussian-shaped pulse propagating in the HCF, with self-steepening effect and higher-order dispersion turned off.** The capillary loss is set to a wavelength-independent constant and the fibre length is set to 50 cm, all other simulation parameters are the same as Supplementary Fig. S5.

# Section 4: Pulse characterization measurements

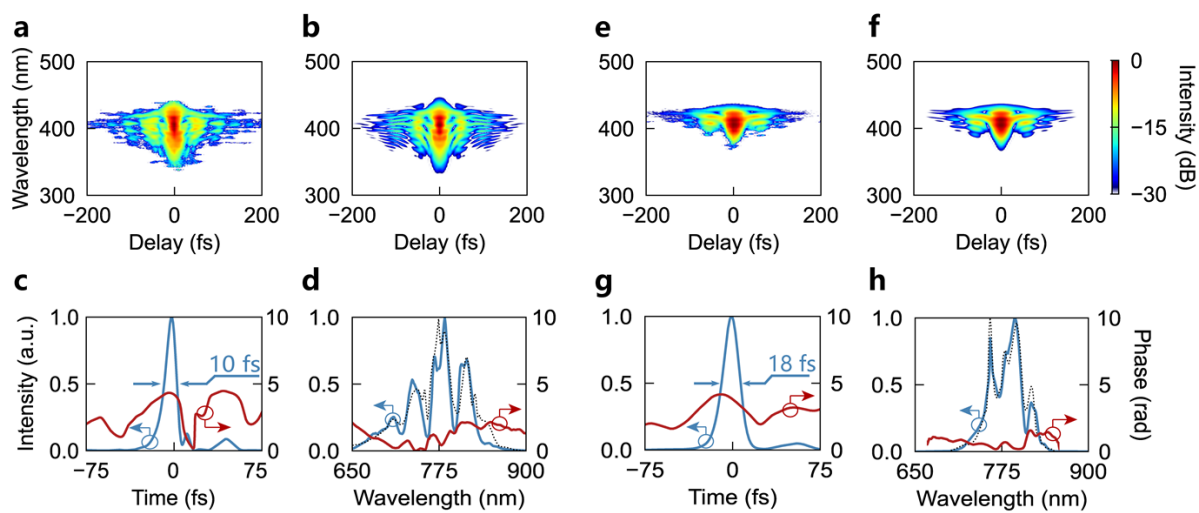

**Supplementary Fig. S10. Characterization of pump pulses using the SHG-FROG set-up.** Measured (a,e) and retrieved (b,f) FROG traces. (c,g) Retrieved temporal (blue line) and phase (red line) profiles with a retrieving error of 0.3%. (d,h) Retrieved spectral (blue line) and phase (red line) profiles. The pump pulse was measured to be 10 fs (a-d) at the gas pressure of 330 mbar and 18 fs (e-h) at the gas pressure of 130 mbar. The experimental spectra are plotted as black dashed lines, measured using a fibre-coupled spectrometer.

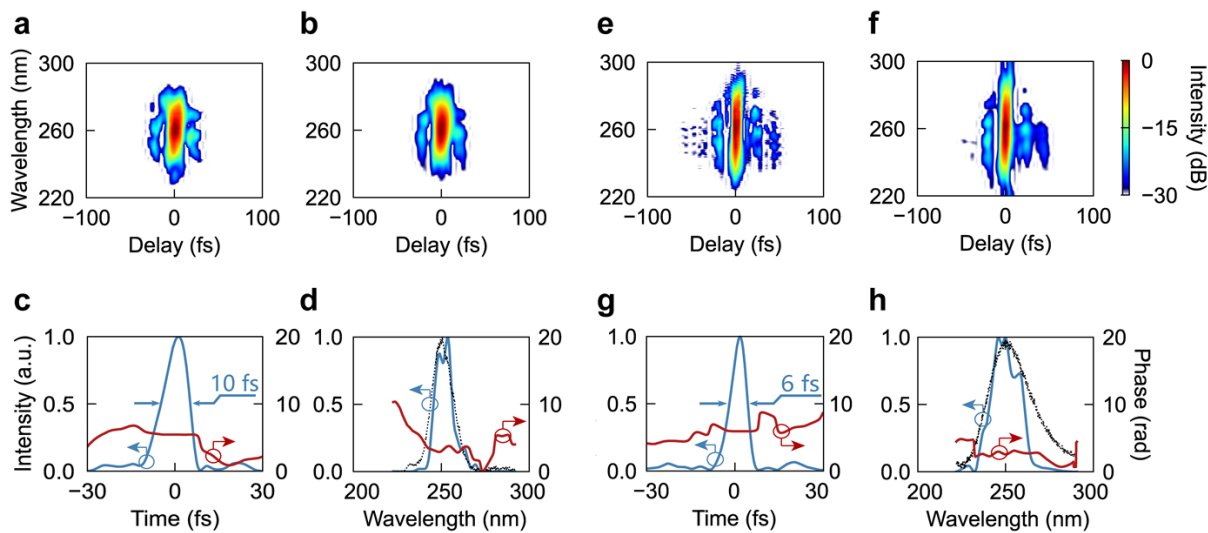

**Supplementary Fig. S11. Characterization of UV DW pulses using the SD-FROG set-up.** Measured (a,e) and retrieved (b,f) FROG traces. (c,g) Retrieved temporal (blue line) and phase (red line) profiles with a retrieving error of 0.5%. (d,h) Retrieved spectral (blue line) and phase (red line) profiles. (a-d) Corresponding to the UV DW pulse in Fig. 1b of the main text, and the pulse was measured to be 10 fs. (e-h) Corresponding to the UV DW pulse in Fig. 3c of the main text when pump pulse energy at 48 µJ, and the pulse was measured to be 6 fs. The experimental spectra are plotted as black dashed lines, measured using a fibre-coupled spectrometer.

# Section 5: Wavelength tunability

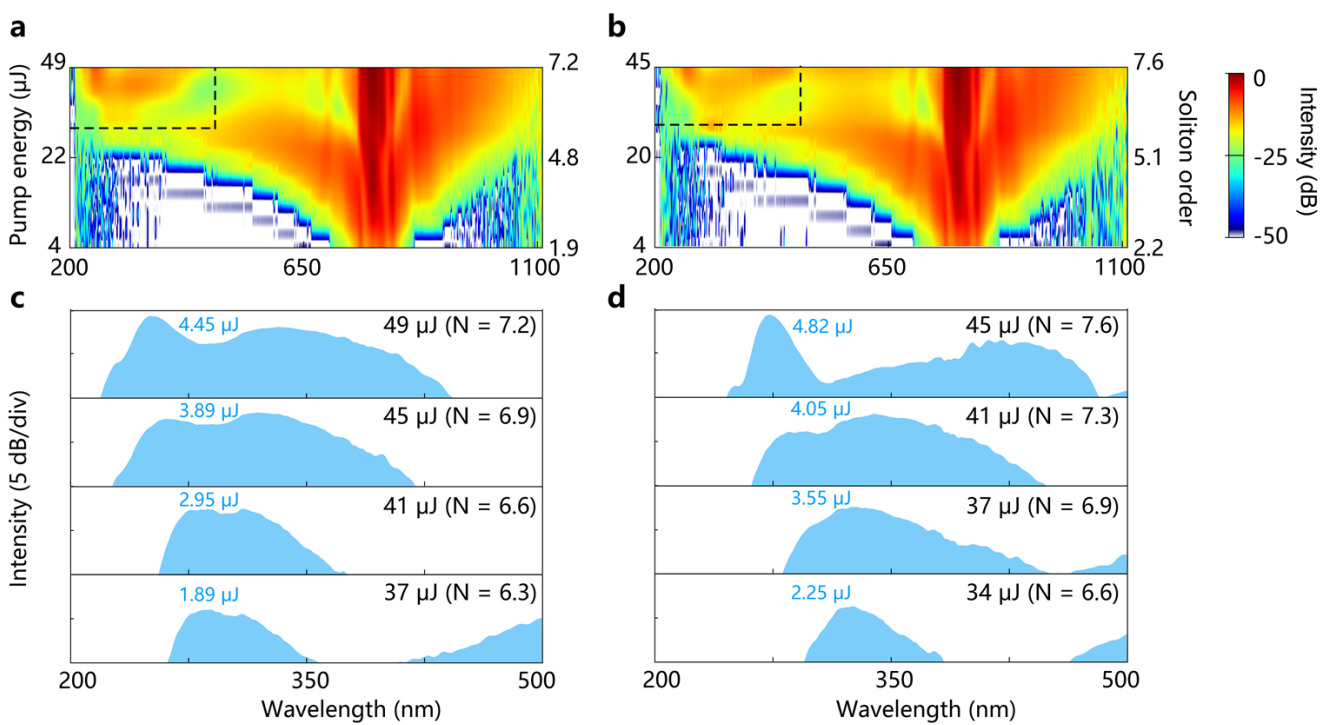

**Supplementary Fig. S12. Experimental results of DW spectral bandwidth adjustments at different gas pressures.** Experimental spectral evolution at the output of the 42-cm-long HCF with 100 μm core diameter filled with 6 bar (a) and 7 bar (b) Ne gas for 18 fs pump pulses as a function of pump pulse energy. (c,d) Several examples of experimental spectral evolution of DW at different pump pulse energies, corresponding to the black dashed frames in (a,b).

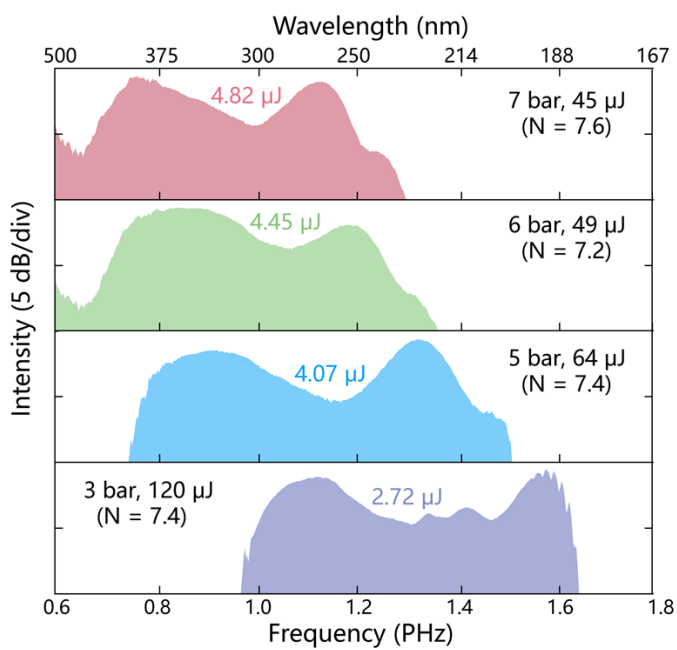

**Supplementary Fig. S13. Central-wavelength tuning of the broadband UV DW.** The broadband DW spectra with different emission wavelengths are measured at different Ne-gas pressure from 3 bar to 7 bar. The pump pulse used in the experiment

has a temporal width of 18 fs and the HCF has a length of 42 cm and a core diameter of 100 µm. These results are plotted on the coordinate axis of spectrum versus frequency.

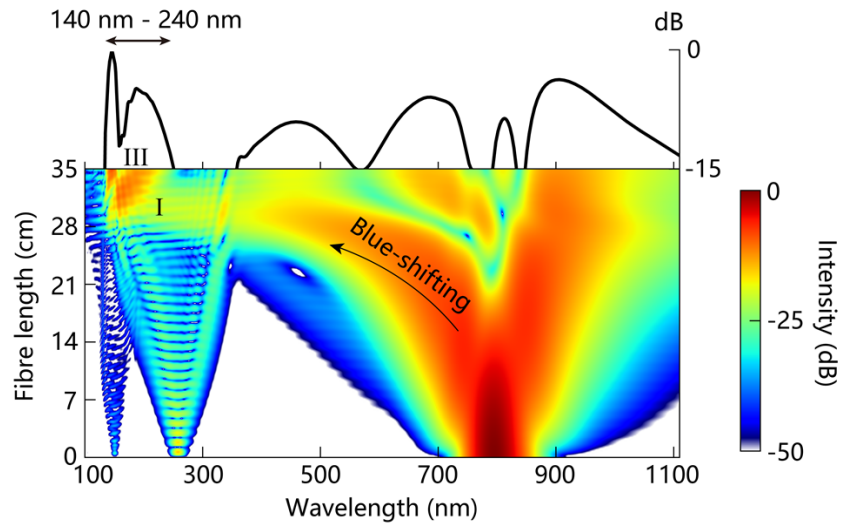

**Supplementary Fig. S14. Broadband DW generation in the vacuum UV regime.** Simulated spectral evolution of 800 nm, 18 fs, 210 µJ Gaussian-shaped pulse propagating in a 35-cm-long, 100-µm-inner-diameter HCF filled with 3.2 bar He gas (N = 7.4) as a function of fibre length. **I** – Narrow-band DW emission, **III** – spectral broadening of the UV DW through cross-phase modulation. The pulse spectrum at the output port of the HCF is plotted as black line.

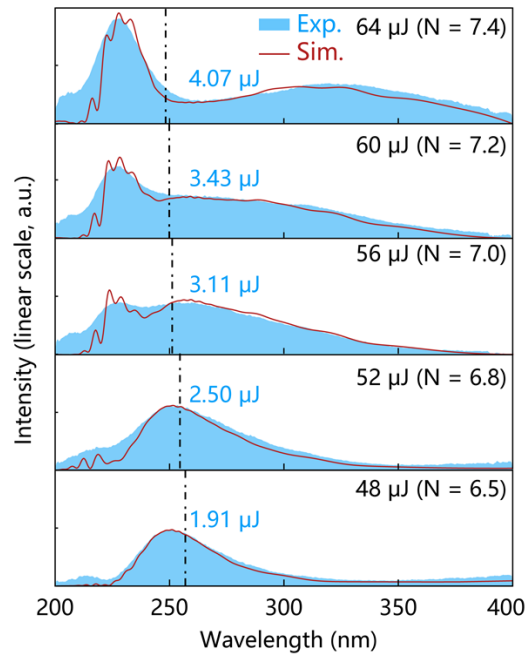

**Supplementary Fig. S15.** Several examples of experimental (blue shadows) and simulated (red lines) spectral evolution of DW at pump pulse energy from 48 µJ to 64 µJ, corresponding to the purple dashed frames in Figs. 3a,b of the main text.

## Supplementary References